\documentclass[traditabstract]{aa}
\usepackage{graphicx}
\usepackage[varg]{txfonts}
\usepackage{natbib}
\usepackage{fixltx2e}
\bibpunct{(}{)}{;}{a}{}{,} 

\usepackage[usenames]{color}
\usepackage{hyperref}
\hypersetup{breaklinks,colorlinks,urlcolor=Blue,citecolor=PineGreen}
\usepackage[hyphenbreaks]{breakurl} 

\title{On-sky characterisation of the VISTA
       NB118 narrow-band filters at 1.19\,$\mu$m\thanks{%
  Based on observations collected at
  the European Southern Observatory, Chile,
  as part of program 284.A-5026 (VISTA NB118 GTO, PI Fynbo) and
  179.A-2005 (UltraVISTA, PIs Dunlop, Franx, Fynbo, \& Le F{\`e}vre)
  }
}

\author{%
Bo Milvang-Jensen\inst{1,2},
Wolfram Freudling\inst{2},
Johannes Zabl\inst{1},
Johan P. U. Fynbo\inst{1},
Palle M{\o}ller\inst{2},
Kim K. Nilsson\inst{2,1},
Henry Joy McCracken\inst{3},
Jens Hjorth\inst{1},
Olivier Le F{\`e}vre\inst{4},
Lidia Tasca\inst{4},
James S. Dunlop\inst{5},
David Sobral\inst{6}
}

\offprints{B. Milvang-Jensen, milvang@dark-cosmology.dk}

\institute{
Dark Cosmology Centre, Niels Bohr Institute,
University of Copenhagen,
Juliane Maries Vej 30, 2100 Copenhagen {\O}, Denmark
\and
European Southern Observatory, Karl-Schwarzschild-Strasse 2,
85748 Garching bei M{\"u}nchen, Germany
\and
TERAPIX / Institut d'Astrophysique de Paris, UMR 7095 CNRS,
Universit{\'e} Pierre et Marie Curie, 98bis Boulevard Arago,
75014 Paris, France
\and
Aix Marseille Universit{\'e}, CNRS,
LAM (Laboratoire d'Astrophysique de Marseille) UMR 7326,
13388 Marseille, France
\and
Scottish Universities Physics Alliance (SUPA),
Institute for Astronomy, University of Edinburgh, Royal Observatory,
Edinburgh EH9 3HJ, UK
\and
Leiden Observatory, Leiden University, PO Box 9513, NL-2300 RA Leiden,
the Netherlands
}

\date{Accepted 2013 October 18
(originally submitted 2013 May 01; resubmitted 2013 October 17)}

\authorrunning{Milvang-Jensen et al.}

\begin{document}

\abstract{%
Observations of the high redshift Universe through narrow-band
filters have proven very successful in the last decade. 
The 4-meter VISTA telescope, equipped with the wide-field camera VIRCAM,
offers a major step forward in wide-field near-infrared imaging, and
in order to utilise VISTA's large field-of-view and sensitivity, the
Dark Cosmology Centre provided a set of 16 narrow-band filters for VIRCAM\@.
These NB118 filters are centered at a wavelength near 1.19$\,\mu$m in a
region with few airglow emission lines. The filters allow the detection of
H$\alpha$ emitters at $z = 0.8$,
H$\beta$ and [\ion{O}{iii}] emitters at $z \approx 1.4$,
[\ion{O}{ii}] emitters at $z = 2.2$, and 
Ly$\alpha$ emitters at $z = 8.8$.
Based on guaranteed time observations of the COSMOS field we here present a
detailed description and characterization of the filters and their
performance. In particular we provide sky-brightness levels and depths
for each of the 16 detector/filter sets and find that some of the filters
show signs of some red-leak. We identify a sample of $2 \times 10^3$
candidate emission-line objects in the data.
Cross-correlating this sample with a large set of galaxies with known
spectroscopic redshifts we determine the ``in situ''
passbands of the filters and find that they are shifted by about
3.5--4\,nm (corresponding to 30\% of the filter width)
to the red compared to the expectation based on the laboratory measurements.
Finally, we present an algorithm to mask out persistence in VIRCAM data.
Scientific results extracted from the data will be
presented separately.
}


\maketitle

\section{Introduction}
\label{sec:Introduction}

The potential of narrow-band searches for redshifted emission lines from
star-forming galaxies has been discussed in the literature for more than two
decades \citep[e.g.,][]{Pritchet_Hartwick:1987, Smith_etal:1989,
Moller_Warren:1993}. With the advent of sensitive detectors on large
telescopes, large samples of e.g.\ Lyman-$\alpha$ (Ly$\alpha$)
emitting objects have been collected
\citep[e.g.,][]{Hu_etal:1998,Kudritzki_etal:2000,Steidel_etal:2000,
Fynbo_etal:2001b,Malhotra_Rhoads:2002, Fynbo_etal:2003b, Hayashino_etal:2004,
Venemans_etal:2005, Kashikawa_etal:2006,
Grove_etal:2009, Ouchi_etal:2009, Ouchi_etal:2010}.
This selection method
combines narrow-band imaging with observations in one or more broad-band
filters. Objects that show excess emission in the narrow-band image compared to
the broad-band images are selected as candidates. The result is a list of
candidate emission-line galaxies within a narrow redshift range, typically
$\Delta z = 0.02 - 0.05$.  
Multi-band photometry can be used to determine the approximate redshift
of the emission line, allowing one to distinguish e.g.\ between
H$\alpha$+[\ion{N}{ii}] on the one hand and
H$\beta$+[\ion{O}{iii}] on the other,
while
spectroscopic follow-up is often necessary to
establish the exact nature of the emission line source
and to measure the precise redshifts.

Of particular interest is the search for Ly$\alpha$
emitters at very high redshift as a probe of the epoch of reionization
\citep{Partridge_Peebles:1967, Barton_etal:2004,Nilsson_etal:2007}. At
redshifts $z>7$, this search requires narrow-band imaging in the
near-infrared (NIR) as the
emission line moves out of the sensitivity range of classical CCDs. Searches
for emission line galaxies based on NIR narrow-band imaging are already
maturing \citep[e.g.,][]{Willis_Courbin:2005,Finn_etal:2005,Cuby_etal:2007,
Geach_etal:2008,Villar_etal:2008,Sobral_etal:2009:Ha,Bayliss_etal:2011,
Ly_etal:2011,Lee_etal:2012,Sobral_etal:2012,Sobral_etal:2013}
but so far we lack any detections of narrow-band selected Ly$\alpha$ emitters
at $z \ga 7.5$ 
\citep[see][and references therein]{Shibuya_etal:2012,Rhoads_etal:2012,
Clement_etal:2012}.


The advent of the VISTA telescope
\citep[e.g.][]{Emerson_etal:2006:Messenger,
Emerson_Sutherland:2010:Messenger,
Emerson_Sutherland:2010}
and its wide-field camera VIRCAM 
\citep{Dalton_etal:2006,Dalton_etal:2010}
provides a new opportunity to undertake a deep,
wide-field search for emission-line galaxies.
To take advantage of this opportunity, we acquired a set of
narrow-band filters (named NB118) for the VISTA telescope.  The filters were
designed to be about 10\,nm wide and centred at around 1185\,nm where there is
a prominent gap in the night sky OH forest \citep{Barton_etal:2004}. 

The NB118 filters allow a search for a number of  line emitters at various
redshifts. The most prominent are H$\alpha$ emitters at $z = 0.8$, H$\beta$ and
[OIII] emitters at  $z\approx1.4$ and [OII] emitters at $z = 2.2$.  The
forbidden oxygen lines are metallicity dependent, but also affected by
active galactic nuclei (AGN)\@.
Nevertheless, in particular [OII] is still a good tracer of star formation and
hence we will have an interesting handle on the star formation density at $z =
2.2$ \citep[e.g.,][]{Sobral_etal:2012,Ly_etal:2012b,Hayashi_etal:2013}, 
which is complementary to
broad- or narrow-band surveys targeting similar redshifts
\citep{Adelberger_etal:2004,Nilsson_etal:2009a}.  Follow-up spectroscopy of
these candidates can provide insight into
the metallicity evolution of
star forming galaxies with redshift \citep{Kewley_Ellison:2008}.
If  sufficient sensitivity can be reached, the filters will also allow a
search for Ly$\alpha$ emitters at $z=8.8$.
This type of survey is currently being undertaken as part of the
ongoing UltraVISTA survey \citep{McCracken_etal:2012}\footnote{%
The narrow-band component of the UltraVISTA survey was originally foreseen
as a stand-alone survey called ELVIS
\citep[Emission Line galaxies with VISTA Survey, e.g.][]{Nilsson_etal:2007}.}. 

In return for providing ESO with the set of 16 NB118 filters
(one per detector), we were awarded 3 nights of
Guaranteed Time Observations (GTO) on VISTA (PI: Fynbo),
cf.\ Sect.~\ref{sec:obs}.
This paper is the first to report results obtained with the NB118
filters on VISTA\@. Therefore, the primary purpose of the paper is
to describe the NB118 filters, to characterize the data
obtained with them, and to describe how best to reduce the
narrow-band images.
In order to quantify the performance of the filters
we report on some preliminary results, but the final scientific
exploitation of the NB118 GTO data will be the subject of a
separate paper.

This paper is organised as follows:
In Sect.~\ref{sec:field_sel_and_obs} and~\ref{sec:datareduc} we describe
our observations and data reduction.
In Sect.~\ref{sec:NB118_filters} we describe the NB118 filters and
predict their filter curves based on laboratory measurements.
In Sect.~\ref{sec:Nbexcess_results} we select objects with narrow-band excess,
cross correlate with spectroscopic redshift catalogues, and
infer the on-sky filter curves.
In Sect.~\ref{sec:NB118_sky_brightness} we analyse the NB118 sky brightness
based on observations, and we investigate indications of red-leaks.
In Sect.~\ref{sec:predicted_sky_brightness} we use models of the sky background
to investigate both the absolute sky brightness and the changes resulting
from a shift of the wavelengths of the filters.
In Sect.~\ref{sec:summary} we summarize the main findings. 
In Appendix~\ref{ap:persistence} we describe the persistence masking
algorithm we developed.

The photometry is on the AB system \citep{Oke_Gunn:1983} and the wavelengths
are in vacuum, unless stated otherwise.

\section{Field selection and observations}
\label{sec:field_sel_and_obs}

\subsection{Field selection}

\begin{figure}[htbp]
\centering
\includegraphics[scale=0.55,bb=19 200 375 492,clip=]{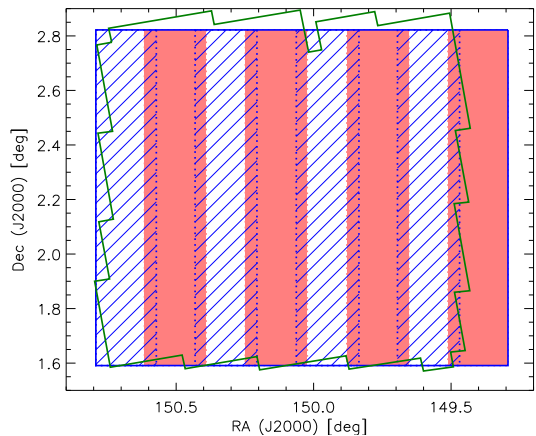}
\caption[]{%
Schematic sky coverage.
Blue hatched columns: stripes observed in this GTO work.
Red filled columns: UltraVISTA ultra-deep stripes.
Blue outline: UltraVISTA contiguous region.
Green outline: HST/ACS region.
\label{fig:fov}
}
\end{figure}

For the GTO programme described in this paper
we targeted part of the COSMOS field \citep{Scoville_etal:2007:COSMOS_intro}
due to its wealth of multi-wavelength data,
including data from the UltraVISTA survey \citep{McCracken_etal:2012}.
We selected our subfield within the COSMOS field in coordination with
UltraVISTA as illustrated by Fig.~\ref{fig:fov}.
The outer blue rectangle of size $\approx 1.5^\circ \times 1.2^\circ$
is the UltraVISTA contiguous region, where UltraVISTA provides imaging
in $Y$, $J$, $H$ and $K_\mathrm{s}$ to varying depths
(referred to as either ``deep'' or ``ultra-deep'').
The 4 filled stripes are the UltraVISTA ultra-deep stripes,
where UltraVISTA additionally provides imaging in NB118.
The 4 hatched stripes are the stripes observed in this GTO program
in NB118 (and to a smaller extent also in $J$);
these stripes have imaging in $Y$, $J$, $H$ and $K_\mathrm{s}$ but not NB118
from UltraVISTA\@.
The area of the stripes is $1\,\mathrm{deg}^2$
(see Sect.~\ref{sec:masking_of_bad_regions}).
The green, jagged outline shows the HST/ACS imaging
\citep{Scoville_etal:2007:COSMOS_HSTACS},
which covers almost the full GTO area.

\subsection{Observations}
\label{sec:obs}

Observations using VIRCAM on VISTA were obtained
in visitor mode during 6 half-nights
(second half of the night),
starting on the night beginning 2010 January 17.
In the first 4 half-nights about 3 hours were spent on NB118 observations
followed by about 1.5 hours on $J$-band observations,
while in the last 2 half-nights all time was spent on NB118 observations.
This is illustrated by
Fig.~\ref{fig:skylevel_vs_time} below, which shows sky level in the
two filters versus time.
The moon was below the horizon all the time.
The seeing in the obtained NB118 images, as computed by the QualityFITS tool,
had a median value of 0.83$''$ and a mean value of 0.89$''$.
The mean airmass of the NB118 observations was 1.19.

\begin{figure*}[htbp]
\centering
\includegraphics[width=1.00\textwidth]{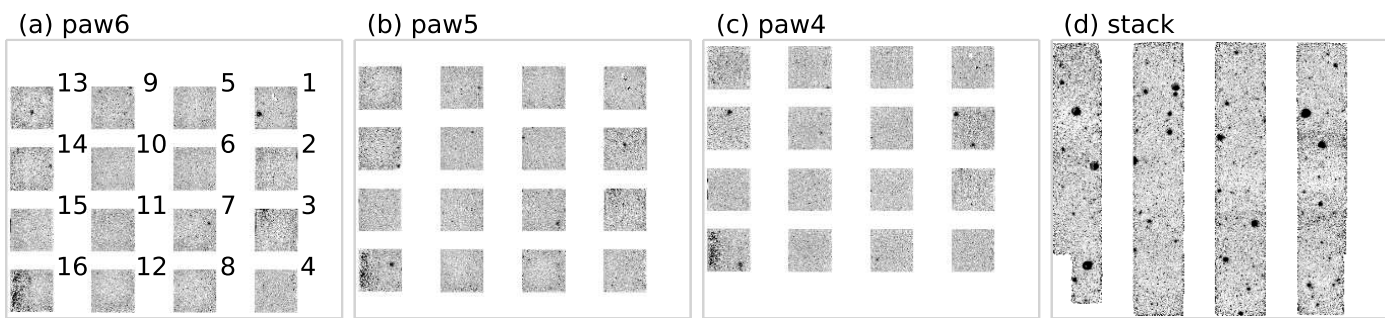}
\caption[]{%
(a)--(c):~Individual VIRCAM NB118 exposures obtained at the 
positions named paw6, paw5 and paw4, respectively.
(d):~The NB118 stack.
In panel~(a) the detector numbers are given.
North is up and east is to the left.
\label{fig:pawprint}
}
\end{figure*}

The sky coverage of VIRCAM in a single exposure, the so-called pawprint,
is illustrated in panels (a)--(c) of Fig.~\ref{fig:pawprint}.
A pawprint covers $0.6\,\mathrm{deg}^2$ on the sky
(16 detectors each $2048^2$ pixels with a scale of $0.34''\,\mathrm{px}^{-1}$).
The 16 detectors are widely spaced, with gaps that are
slightly less than a full detector in X and half a detector in Y\@.
The 3 particular positions on the sky shown by the pawprints in panels (a)--(c)
are named paw6, paw5 and paw4 within the UltraVISTA project.
They are spaced in Y (here Dec) by 5.5$'$, and
by combining exposures taken at these 3 positions, one gets a set of
4 stripes or columns (see Fig.~\ref{fig:pawprint}d) in which each pixel
typically receives data from 2 of the 3 pawprints.

The observing strategy employed in this project was to obtain a single
exposure at each pawprint position and then move to the next,
in the sequence paw6, paw5, paw4; paw6, paw5, paw4; etc. 
The moves paw6$\rightarrow$paw5 and paw5$\rightarrow$paw4 
were always 5.5$'$ in Dec, i.e.\ with no random component,
and only in the paw4$\rightarrow$paw6 move a random component (jitter),
drawn from a box of size $122'' \times 122''$ in RA $\times$ Dec,
was added.\footnote{%
This sequence corresponds to the nesting parameter in the
OBs (Observation Blocks)
being FJPME, with the loop over J (jitter)
being outside the loop over P (pawprint); 
see the discussion in \citet{McCracken_etal:2012}.
The reverse order of J and P had been desirable.
}
The lack of a random component in some of the telescope moves meant that fake
sources (so-called persistent images) were present when we first stacked the
data, despite combining the individual images using sigma clipping.
Only after developing a method to mask the fake sources in the individual
images (see Sect.~\ref{sec:persistence}), the resulting stack was free of such
fake sources.

The exposure times of the individual images were
280~sec in NB118 (NDIT = 1, DIT = 280~sec) and
120~sec in $J$ (NDIT = 4, DIT = 30~sec).
The total exposure time obtained is listed in Table~\ref{tab:total_exposure}.

\begin{table}
\caption{Total usable exposure time from the GTO programme
\label{tab:total_exposure}
}
\centering
\begin{tabular}{lrr}
\hline
                       & NB118 [s] & $J$ [s] \\ \hline
paw6                   & 23800     &  6240  \\
paw5                   & 24080     &  6000  \\
paw4                   & 24080     &  6000  \\ \hline
stack                  & 47973     & 12160  \\ \hline
\end{tabular}
\tablefoot{%
The total usable exposure time is listed, i.e.\
after discounting one NB118 exposure rejected in the visual inspection
(Sect.~\ref{sec:terapix})
and 12 $J$-band exposures not delivered by CASU (Sect.~\ref{sec:CASU}).
The exposure time for each of the 3 partially overlapping pawprint
positions is given (cf.\ Fig.~\ref{fig:pawprint}), as well as the
typical exposure time per pixel in the stack, calculated as
2/3 times the sum over all pawprints.
}
\end{table}

\subsection{Additional imaging}
\label{sec:additional_imaging}

In addition to the NB118 and $J$-band VISTA data obtained in this programme,
we used $Y$ and $J$-band VISTA data from the UltraVISTA DR1 dataset
\citep{McCracken_etal:2012}.
Specifically, for $Y$ we directly use the stack and weight map
from \citet{McCracken_etal:2012},
although we add 0.04\,mag to the zeropoint to reproduce the
effect of the latest photometric calibration (colour equation) from CASU
(Sect.~\ref{sec:CASU}).
For $J$, we use the individual images and weight maps
(which \citealt{McCracken_etal:2012} used to make their stack)
and combine them with the individual images and weight maps from
our programme (Sect.~\ref{sec:terapix}).
This combined $J$-band stack has a typical exposure time per pixel of 17.2\,h
in the stripes of interest here,
of which 13.8\,h come from UltraVISTA \citep[Table~2 in][]{McCracken_etal:2012}
and 3.4\,h come from our programme (Table~\ref{tab:total_exposure}).
We use the NB118, $J$ and $Y$ photometry to select candidate emission line
objects (Sect.~\ref{sec:selection_Nbexcess}).

\section{Data reduction}
\label{sec:datareduc}

\subsection{Processing in the Data Acquisition System}
\label{sec:Correlated_Double_Sample}

VIRCAM uses Correlated Double Sample (CDS) readout mode,
also known as Fowler-1 sampling \citep{Fowler_Gatley:1991,McMurtry_etal:2005}
and as a Reset-Read-Read sequence.
This means that if the user requests a single 280 seconds exposure
(DIT = 280 seconds, NDIT = 1),
the Data Acquisition System (DAS) will (a)~reset the detectors,
(b)~integrate on sky for 1 second and read,
(c)~integrate on sky for 281 seconds and read, and
(d)~write the difference between the second and the first readout
to disk.
(For $\mathrm{NDIT} > 1$ the above procedure is carried out NDIT times,
and the delivered FITS file contains the sum of the NDIT image differences.)
This process is transparent to the user, but it has implications for
the appearance of saturated objects (Sect.~\ref{sec:persistence}),
as well as for estimating the linearity of the detectors
(Sect.~\ref{sec:CASU} and \citealt{Lewis_etal:2010}).
It also means that the pedestal (bias) level set by the readout electronics
has already been subtracted by the DAS, removing the need for such a
reset correction in the subsequent reduction pipeline (Sect.~\ref{sec:CASU}).
The effective readout noise resulting from this readout mode is about
$23\,e^-$ on average over the 16 detectors
(see e.g.\
the VIRCAM/VISTA User Manual).
No other readout modes are offered.

\subsection{Processing at CASU}
\label{sec:CASU}

All VISTA data are processed by the VISTA Data Flow System (VDFS)
\citep[e.g.][]{Emerson_etal:2004}, which consists of
quality control pipelines at ESO Paranal and Garching
\citep[e.g.][and the VIRCAM/VISTA User Manual]{Hummel_etal:2010},
a 
science reduction pipeline at the Cambridge Astronomy Survey Unit (CASU)
\citep[e.g.][and the CASU web site\footnote{\url{http://casu.ast.cam.ac.uk/surveys-projects/vista/technical/data-processing}}]{Irwin_etal:2004,Lewis_etal:2010},
and a generation of futher data products and archiving
at the VISTA Science Archive (VSA) at
the Wide-Field Astronomy Unit (WFAU) in Edinburgh
\citep[e.g.][]{Hambly_etal:2004,Cross_etal:2012}.
For the present work, only the CASU processing is relevant.
The `raw' (as coming from the DAS, see
Sect.~\ref{sec:Correlated_Double_Sample})
individual images, i.e.\ 
the set of 16 detector images for a single exposure,
undergo the following processing steps in the CASU science pipeline:
\begin{itemize}
\item
Dark correction, performed by subtracting a combined dark image based on
individual dark images with
the same DIT and NDIT as the science image in question; this corrects
both for the thermal dark current and for the effect termed
`reset anomaly'
\citep[e.g.][and the VIRCAM/VISTA User Manual]{Irwin_etal:2004}.
\item
Nonlinearity correction, derived from screen flats (dome flats)
of different exposure times, taken under a constant light level.
\item
Flat field correction,
performed by division by a normalised, combined twilight sky flat field
image.
This step corrects for small-scale variations in the
quantum efficiency within each detector
as well as for the vignetting of the camera.
This step also corrects for more effects by virtue of the normalisation:
all 16 detector images of the flat field are normalised by the same number,
namely the mean (over the 16 detectors) of the median level in each detector.
The resulting flat does not have a level near 1 in each detector; rather,
it has a level of about 1.15 in detector 1 and 0.70 in detector 2,
for example (this value is recorded in the GAINCOR header keyword for the
given detector).
This means that the combined effect of detector-to-detector differences
in gain (in $e^-\,\mathrm{ADU}^{-1}$)
[ADU: analogue-to-digital unit] and in overall quantum efficiency (QE)
are removed. The unit of the counts in the flat-fielded science images
is termed `gain-normalised ADU'\@. If QE differences are ignored,
these counts can indeed be converted to electrons using
a single gain valid for all detectors; this gain is about
4.2 electrons per gain-normalised ADU.
\item
Sky background correction or sky subtraction, performed by subtracting
a sky image. The sky images are made by splitting the time sequence of science
images into blocks; for this dataset blocks of 6 images were used.
The objects in the images are masked (for this dataset using a mask we
provided, which was based on existing $K_\mathrm{s}$ and $i$-band data), and
then these 6 object-masked images are combined to form the sky frame. 
This single sky frame is normalised by subtracting its median level
and then subtracted from the 6 science images.
\item
Destriping, which removes stripes caused by the readout electronics.
The stripes are horisontal in the detector $x,y$ coordinate system
contained in the raw FITS files.
For our data taken with a position angle of $0^\circ$, the stripes are
vertical in our astrometrically calibrated images having north up and
east left.
\item
Astrometric calibration,
based on the 2MASS catalogue \citep{Skrutskie_etal:2006}.
\end{itemize}
The VIRCAM data show no evidence of detector crosstalk or sky fringing,
so the pipeline does not need to correct for these effects
\citep{Lewis_etal:2010}.

The reduced individual images for each OB and each pawprint are stacked
(hereafter referred to as the \texttt{\_st} stacks).
For example, our NB118 data resulted in 18 \texttt{\_st} stacks, since our OBs
obtained data at 3 pawprint positions and since one OB was executed per night
for 6 nights.
In these stacks, a photometric calibration onto the
VISTA photometric system is performed, see
the CASU web site\footnote{\url{http://casu.ast.cam.ac.uk/surveys-projects/vista/technical/photometric-properties}},
the presentation by S. Hodgkin\footnote{\url{http://casu.ast.cam.ac.uk/documents/vista-pi-meeting-january-2010/VISTA-PI-sth-calib.pdf}},
and the paper by \citet{Hodgkin_etal:2009} describing the analogous
calibration for the UKIRT/WFCAM photometric system.
The calibration works as follows.
One the one hand instrumental total magnitudes 
for stars in the image
are calculated from the flux (in units of gain-normalised ADU per second),
corrected for the radially changing pixel size.
On the other hand magnitudes in the VISTA photometric system (Vega) are
predicted based on the 2MASS $J$, $H$ and $K_\mathrm{s}$ (Vega) magnitudes
for stars in the image 
using colour equations, which for $Y$, NB118 and $J$ are
\begin{eqnarray}
       {Y}_\mathrm{VISTA,predicted} & = &
  J_\mathrm{2MASS} + 0.550 \cdot (J-H)_\mathrm{2MASS} \label{eq:Y_pred} \\
\mathrm{NB118}_\mathrm{VISTA,predicted} & = &
  J_\mathrm{2MASS} + 0.100 \cdot (J-H)_\mathrm{2MASS} \label{eq:NB118_pred} \\
       {J}_\mathrm{VISTA,predicted} & = &
  J_\mathrm{2MASS} - 0.070 \cdot (J-H)_\mathrm{2MASS} \label{eq:J_pred}
\end{eqnarray}
and where the stated coefficients refer to version 1.0 of
the CASU VIRCAM pipeline\footnote{\url{http://casu.ast.cam.ac.uk/surveys-projects/vista/data-processing/version-log}} which applies to the data used here;
from CASU version 1.1, the coefficient 
for $Y$ was changed to 0.610 and for $J$ to $-0.077$.
The advantage of using standard stars (here the 2MASS stars) located in the
image itself is that a zeropoint can be calculated simply by comparing the
predicted magnitudes with the instrumental magnitudes ---
the actual atmospheric extinction for the given image (even including
possible clouds) is automatically included.
However, the CASU pipeline calculates a zeropoint `corrected' to airmass
unity, which makes it easier to monitor e.g.\ the instrument throughput,
but which mandates a reverse correction when users want to
transform instrumental magnitudes into magnitudes on the
VISTA photometric system (cf.\ appendix~C in \citealt{Hodgkin_etal:2009}).
The equation for the CASU zeropoint, here for the $Y$ band, reads
\begin{equation}
\mathrm{ZP}(Y) = \mathrm{median} \left\{
  {Y}_\mathrm{VISTA,predicted} - Y_\mathrm{instrumental}
\right\} + k \, (X-1)
\enspace ,
\label{eq:ZP_Y}
\end{equation}
where the median is taken over the used stars
(typically 380 stars per image for our NB118 data), $X$ is the airmass, and
$k$ is the atmospheric extinction coefficient, which seems to be
0.05\,mag/airmass for all bands (listed in header keyword EXTINCT)\@.
This zeropoint at airmass unity is written to the header (keyword MAGZPT)\@.
For our NB118 and $J$-band data, the value of this keyword was the same
for all 16 detectors. It makes sense that the zeropoint is almost
the same for all detectors since the flat fielding has removed
detector-to-detector differences in both gain and in QE;
however, the colour of the twilight sky (used in the flat fielding) and
of the astronomical objects of interest may differ, and filter-to-filter
differences (since each detector has its own filter) could also be
relevant.
A robust estimate of the standard deviation of the differences between
predicted and instrumental magnitudes over the used stars is written
to keyword MAGZRR, which typically was 0.018\,mag for our NB118 data.

For completeness it should be mentioned that, as written here, the
equations for the predicted magnitudes (Eq.~\ref{eq:Y_pred}--\ref{eq:J_pred}),
or equivalently the equation for the zeropoint (Eq.~\ref{eq:ZP_Y}),
miss a term of the form $-c \, E(B-V)'$,
where $E(B-V)'$ is calculated from the Galactic reddening $E(B-V)$
from \citet{Schlegel_etal:1998} using Eq.~1 in \citet{Bonifacio_etal:2000}.
This term corrects for the different stellar population mix found in highly
reddened parts of the sky. 
The constant $c$ is generally small, e.g.\ 0.14 for $Y$,
and it is very small for bands within the $JHK_\mathrm{s}$
wavelength domain of 2MASS, e.g.\ 0.01 for $J$.
The term does not correct for Galactic extinction as such.

Processed data were made available to us by CASU on 2010 July 21.
%
This included reduced individual images, calibration frames
(darks, flats and sky frames) and \texttt{\_st} stacks. We process these
data further, as described in Sect.~\ref{sec:terapix}.
The essence is that we undo the CASU sky subtraction in the individual
images, apply our own sky subtraction and stack these images,
masking fake sources (``persistent images'', see Sect.~\ref{sec:persistence})
at the same time.
We do not use the \texttt{\_st} stacks, except that we use the
photometric calibration contained in their headers.

Reduced individual images were provided for all 258 obtained NB118 exposures,
but only for 152 of the 164 obtained $J$-band exposures.
The latter problem seems to be due to the $J$-band OBs 
being interrupted and restarted at two points due to minor technical (software)
problems, which 
due to an unfortunate interplay between ESO's automatic image grading and
the CASU pipeline made some images be rejected.
Since plenty of additional $J$-band data subsequently became available
from UltraVISTA, this minor loss of data was not a problem.

\subsection{Creation of persistence masks}
\label{sec:persistence}

The VIRCAM detectors do have some persistence, in which a somewhat bright star
(down to say $J \approx 16$) leaves a faint fake source at the same
position on the detector in the next 1 or 2 images.
For our observing pattern (nesting) such faint fake sources would add up
in the stack, and it was therefore necessary to deal with this problem.
We developed a masking algorithm that masks the affected
pixels in the individual images, thus excluding those pixels when stacking
the data. This algorithm is described in Appendix~\ref{ap:persistence}.
An illustration of persistence in individual VIRCAM images is given
in the top row of Fig.~\ref{fig:persistence_indiv_images}.
The effect of persistence in the stack of our data without and with
our persistence masking is illustrated in
Fig.~\ref{fig:persistence_stack}.

\subsection{Processing at TERAPIX}
\label{sec:terapix}

We used the TERAPIX (Traitement {\'E}l{\'e}men\-taire, R{\'e}duction et
Analyse des PIXels) facility to process the individual reduced images from
CASU (Sect.~\ref{sec:CASU}).
This processing was similar to that done for the UltraVISTA DR1 data,
and we refer to \citet{McCracken_etal:2012} for details.
Here we will mention the main points.

The 410 individual reduced images ($\times$ 16 detectors) from CASU
(258 NB118 and 152 $J$-band images)
and a number of diagnostic plots based on these were visually inspected 
within the Youpi\footnote{\url{http://youpi.terapix.fr/}} environment
\citep{Monnerville_Semah:2010}. 
One NB118 image, namely the very first taken for this project,
was found to contain a strange diagonal stripe,
and was therefore flagged for exclusion from the stacking.
Unlike for UltraVISTA DR1, we did not additionally reject images based
on the PSF size or ellipticity.

For each image, a weight map was constructed as a copy of
the flat field from CASU\@. Pixels in the bad pixel mask from CASU were
set to zero. At this point all the weight maps would be identical and
resemble the confidence map from CASU, which we do not use.
For each weight map, a value of zero was assigned to the pixels in the
computed persistence mask (Sect.~\ref{sec:persistence}) for the given image,
enabling persistence-free stacks to be made.
The weight maps were generated using the
WeightWatcher tool \citep{Marmo_Bertin:2008}.
The weight maps are used in the sky subtraction and in the stacking.

As the first step in the two-step sky-subtraction procedure,
the individual images from CASU were stacked to produce a first-pass stack.
From this, an object mask was generated and subsequently
transformed (resampled) to create an object mask for each individual image.

As the second step, we apply our own sky subtraction.
We first undo the CASU sky subtraction. Specifically, to each individual
reduced image from CASU we add the sky frame from CASU for that image.
A new sky subtraction was performed with the following three key features:
(1)~Masking of objects was based on the above-mentioned first-pass stack of
all the data
plus a mask generated from the objects detected in the image itself
(thus including artefacts not found in the stack, e.g.\ satellite trails
and cosmics).
For comparison, the CASU sky subtraction for this dataset
used masking based on a mask we provided, which was based on existing
$K_\mathrm{s}$ and $i$-band data.
(2)~For each image a sky frame was constructed from the images
(excluding the image itself)
in a window of $\pm20\,\textrm{min}$ centered on the image in question
(also known as running sky or sliding sky). Since the NB118 images typically are
spaced by 4.9\,min, typically 8 images were used to create the sky frame,
namely 4 images before and 4 after.
For the images at the start or end of a sequence (an OB),
the window was made asymmetric so that it would still contain
8 images (this differs from the UltraVISTA DR1 processing).
For comparison, the CASU sky subtraction for this dataset
consisted of creating a sky frame based on fixed groups of 6 NB118 images
(corresponding to a time span of about 30 minutes) and using that
sky frame to sky-subtract all 6 images. When inspecting the individual images
as sky subtracted by CASU,
we noted that the large-scale sky background was less flat for the
images at the ends of such a 6-image sequence than for the images
in the middle of the sequence.
(3)~Large-scale gradients were fitted and subtracted using
SExtractor\footnote{\url{http://www.astromatic.net/software/sextractor}}
\citep{Bertin_Arnouts:1996}, with the objects being masked using
the object masks also used in the sky subtraction process.
Finally an additional destriping was done.
After the sky subraction, the weight maps were updated,
assigning a value of zero to pixels where no sky frame could be computed;
this happens when the given pixel is masked in all the images used to
create the sky frame.
Also the catalogues needed for SCAMP (see below) were remade.

We note in passing that an earlier version of the
TERAPIX sky-subtraction code had a bug involving
images sometimes are shifted by 1 pixel.
That bug has been fixed in this work.
The bug affected the NB118 UltraVISTA DR1 stack.

For each image the astrometric and photometric solutions were computed using
SCAMP\footnote{\url{http://www.astromatic.net/software/scamp}}
\citep{Bertin:2006}.
The photometric calibration is based on that provided by CASU\@.
As described in Sect.~\ref{sec:CASU}, a photometric zeropoint is only
derived for the \texttt{\_st} stacks, of which there are 18 for the NB118 data.
The zeropoint is listed as corrected to airmass unity, and we undo this
correction, making the zeropoint applicable to the actual airmass.
As the intial guess of the zeropoint for a given individual image we use
the zeropoint of the \texttt{\_st} stack in which the image is part.
Using a catalogue for each image, SCAMP then compares the object fluxes
between images and for each image computes a flux scale factor that
will bring all images to agree. The absolute zeropoint is computed
as the average over all the initial zeropoints:
in the language of SCAMP all images were classified as photometric.
While not all images may be photometric, the CASU zeropoints were computed
from stacks of all the images, and therefore all images should be used
to derive overall zeropoint.
A conversion from Vega to AB was also included, using the same
offsets as \citet{McCracken_etal:2012}.
The astrometric calibration was computed using the
COSMOS CFHT $i$-band astrometric reference catalogue
\citep{Capak_etal:2007,McCracken_etal:2010,McCracken_etal:2012}.
For the NB118 data,
the internal astrometric scatter
was $0.043''$ in RA and $0.062''$ in Dec for about 30,000 high S/N objects.
The external astrometric scatter
was $0.075''$ in RA and $0.077''$ in Dec for about 400 high S/N objects.
%
%
%

The individual images were regridded and stacked using
SWarp\footnote{\url{http://www.astromatic.net/software/swarp}}
\citep{Bertin_etal:2002}.
The regridding (interpolation) was done to the same tangent point and
pixel scale of  $0.15''\,\mathrm{px}^{-1}$ used in UltraVISTA and in
most available images for the COSMOS field. For reference, the native pixel
of VIRCAM is $0.34''\,\mathrm{px}^{-1}$.
The stacking was done using sigma clipping (at 2.8$\sigma$);
a modified version of SWarp was used to accomplish this.
The output files from SCAMP were used to define the astrometry and the
photometric zeropoint of the stack.
We created two stacks: an NB118 stack based on the GTO data,
and a $J$-band stack based on both the GTO and the UltraVISTA DR1 data
\citep{McCracken_etal:2012}.

\subsection{Test using the CASU sky subtraction}
\label{sec:stack_CASU_skysubtraction}

As a test, we stacked the individual NB118 images as sky subtracted by CASU\@.
This was done in almost the same way as for the individual images as
sky subtracted by TERAPIX\@. Specifically, SWarp was run with the same
parameters and the same individual SCAMP files (containing the
astrometric and photometric solutions) were used.
The only difference was that we let SWarp fit and subtract
large-scale gradients in the individual images, since such gradients
were clearly present. 

Compared visually to the TERAPIX stack,
the CASU-based stack had slightly more cosmetic problems on large scales.
On small scales, the CASU-based stack sometimes showed stripes, but
otherwise this stack appeared at least as deep as the TERAPIX stack.
The result from empty aperture measurements is given in Sect.~\ref{sec:depth}.

\begin{figure*}[htbp]
\centering
\includegraphics[width=0.98\textwidth,bb=2 529 572 734]{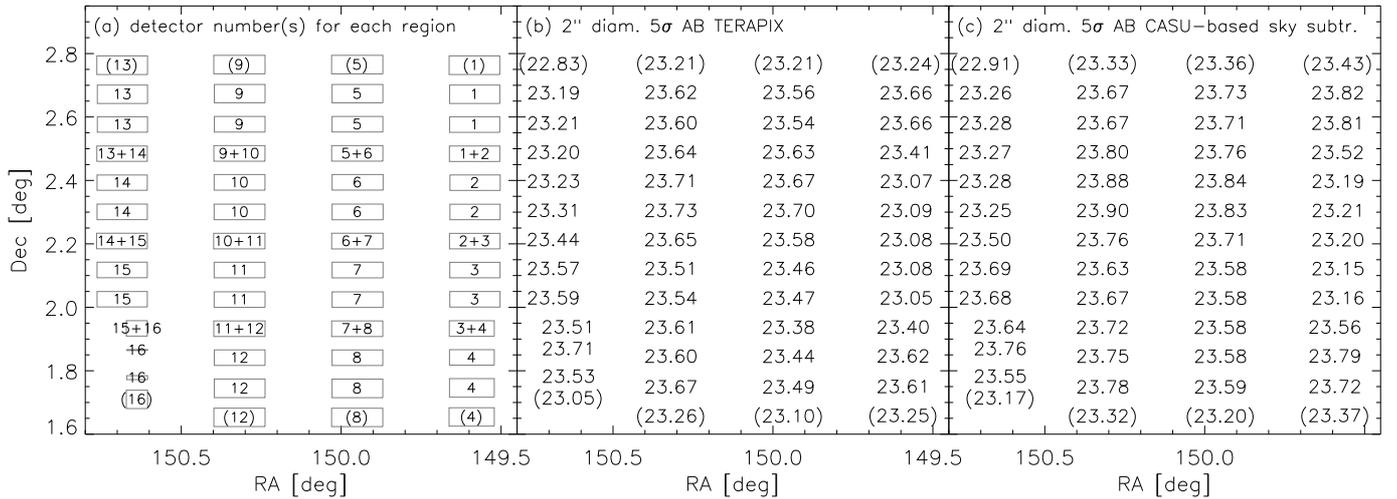}
\caption[]{%
Empty aperture noise measurements for the GTO NB118 data.
The rectangles in panel~(a) show the regions
used to analyse the empty aperture measurements.
The detector(s) contributing data in the given region are listed,
and parenthesis indicate regions that have half the
exposure time of the other regions.
Panels~(b) and (c) give the 2$''$ diameter $5\sigma$ AB mag values
for the NB118 TERAPIX stack (Sect.~\ref{sec:terapix})
and our NB118 test stack based on the CASU sky subtraction (which has 
about 0.1\,mag lower noise,
but which still may not be the best possible version of our data;
Sect.~\ref{sec:stack_CASU_skysubtraction}), respectively.
The aperture correction to total magnitude of 0.34\,mag has not
been subtracted.
\label{fig:depth}
}
\end{figure*}

Note that in the above test, we stacked the 257 \emph{individual} NB118 images
as sky subtracted by CASU
(and we first removed the large-scale gradients),
not the 18 \texttt{\_st} stacks made by CASU (at the native
pixel scale).
%
We expect that (a)~only stacking once, i.e.\ going from the individual images
to the stack, and (b)~using a finer pixel scale to (marginally) recover
spatial resolution is the better procedure.

\subsection{Photometry}
\label{sec:photometry}

Photometry was performed using SExtractor version 2.8.6.
Objects were detected/defined
in the NB118 image, and fluxes in identical apertures were measured
in the NB118, $J$ and $Y$-band images.

Isolated bright but unsaturated stars were located in a
FWHM versus magnitude plot. 
The typical seeing (FWHM, as measured by SExtractor), was
0.89$''$ for NB118,
0.87$''$ for $J$, and
0.88$''$ for $Y$\@.
Using circular apertures of 2.0$''$ and 7.1$''$ diameter
\citep[following][]{McCracken_etal:2012},
the aperture correction between these two apertures was found to be
0.34\,mag for NB118,
0.32\,mag for $J$, and
0.35\,mag for $Y$\@.
Typically, 1000 stars were used and the standard deviation was 0.01\,mag.
The 2.0$''$ diameter aperture magnitudes with these aperture corrections
subtracted are used throughout this paper. It should be noted that the
used aperture corrections only are correct for unresolved objects.
Conversely, it should be noted that the different bands have almost the
same seeing, so the aperture corrections are not critical for the derived
colours.

The errors on the aperture magnitudes computed by SExtractor are too small
due to, among other things, correlated errors introduced by the resampling.
We used the empty aperture measurements (Sect.~\ref{sec:depth}) to derive
a typical correction factor of 2.7 to the SExtractor flux (i.e.\ counts) errors.
This method is similar to the simulations done by \citet{McCracken_etal:2010}.

\subsection{Masking of bad regions of the stacks}
\label{sec:masking_of_bad_regions}

In the analysis we mask certain regions of the stack where
it is difficult to extract correct photometry.
In a zone of width 122$''$ around the edges of the 4 stripes the
exposure time linearly decreases due to the jittering (Sect.~\ref{sec:obs}).
This is mostly not a problem since the weight map of the stack tracks this,
but very close to the edge the photometry is unreliable, as seen by
objects being detected in NB118 but not in $J$, and objects showing
narrow-band excess but having a spectroscopic redshift that does not
match known strong emission lines.
We therefore mask typically 15$''$ around the edges.
We also mask regions contaminated by reflections from bright stars,
also based on a visual inspection.
The area of the stack containing data, defined as pixels
with a positive value in the weight map, is 1.08\,deg$^2$ before masking.
This area includes the regions of height $\approx$5.5$'$ at the top and bottom
of the stack where the exposure time is only half of that in the main part
of the stack, and it includes the 122$''$ around the edges where the
exposure time is lower due to the jittering.
After the masking the area containing data is 0.98\,deg$^2$.

\subsection{Depth of the obtained stacks}
\label{sec:depth}

To measure the depth in the stacks we proceed as follows.
We first run SExtractor in the given stack to detect the objects.
This produces a so-called segmentation image, which identifies all the pixels
that contain signal from objects.
We then place as many non-overlapping circles of 5$''$ diameter as possible
in the image in such a way that the circles do not contain any object pixels.
At the centre of these circles we force SExtractor to perform aperture
photometry in circular apertures with a range of sizes; here we will report
the results for the 2$''$ diameter apertures.

To accurately track the depth as function of detector (or as function of
each of the 16 NB118 filters), we identify regions in the stack that
are fully covered 
by exactly 2 of the 3
pawprints. These regions are shown in Fig.~\ref{fig:depth}(a).
On the figure the detector number(s) contributing data to the given region
are given. A label such as ``2'' indicate that the region only contains
data from detector 2. There are two such regions:
the top one gets data from paw4 and paw5, and
the bottom one from paw5 and paw6.
The area between these two regions is covered either by 2 or 3 pawprints
(the average is 2.2 pawprints) and hence has a 10\% larger exposure time
per pixel. These areas are not analysed in the following depth analysis.
The regions with the detector number given
in parenthesis are special cases: they are fully covered by exactly 1 pawprint
and thus have half the exposure time per pixel of the other regions.

For detector 16 the regions take into account that our weight maps for
the individual images remove 3.3$'$ on the south side and
5.5$'$ on the east side as this part of the detector is deemed unreliable.
For detector 4 we remove 1$'$ on the west side.

Each region contains typically 2,500 empty aperture flux measurements.
The standard deviation $\sigma$ of these values is calculated and
turned into a $5\sigma$ AB magnitude.
The results for our TERAPIX stack (Sect.~\ref{sec:terapix})
are shown in Fig.~\ref{fig:depth}(b) and
for our test stack of the individual images as sky subtracted by CASU
(Sect.~\ref{sec:stack_CASU_skysubtraction}) 
in Fig.~\ref{fig:depth}(c).
It is seen that the $5\sigma$ AB noise typically is about 0.1\,mag
worse in our TERAPIX stack than in the CASU-based stack.
The reason for this is unknown.
It indicates that the reduction can be marginally improved.
A difference in depth at this level does not affect the conclusions
in this paper.

For both stacks there is a substantial variation in the noise within
the stack. Using the numbers from Fig.~\ref{fig:depth}(c),
the lowest noise (ca.\ 23.9\,mag) is found for detectors 1, 6 and 10, and
the highest noise (ca.\ 23.2\,mag) is found for detectors 2, 3, 13 and 14.
This difference strongly tracks the difference in sky brightness,
as will be discussed in Sect.~\ref{sec:NB118_sky_brightness_GTO}
--- see e.g.\ Table~\ref{tab:sky_level}.

Over the 44 regions with full exposure (cf.\ Fig.~\ref{fig:depth}a),
the median $5\sigma$ noise is 23.54\,mag for the TERAPIX stack and
23.67\,mag for the CASU-based stack.
Subtracting the point source aperture correction of 0.34\,mag
(Sect.~\ref{sec:photometry})
we obtain the corresponding median $5\sigma$ detection limits of
23.20\,mag and
23.33\,mag, respectively.
These values can be converted into median $5\sigma$ detection limits
in line flux of
$5.0 \times 10^{-17}\,\mathrm{erg}\,\mathrm{s}^{-1}\,\mathrm{cm}^{-2}$ and
$4.4 \times 10^{-17}\,\mathrm{erg}\,\mathrm{s}^{-1}\,\mathrm{cm}^{-2}$,
respectively, via
\begin{equation}
F = 
3.0 \times 10^{18} \,
\frac{w \, \mathrm{\AA}}
     {\lambda^2} \,
10^{-0.4(m_\mathrm{AB}+48.60)} \,
\mathrm{erg}\,\mathrm{s}^{-1}\,\mathrm{cm}^{-2} \enspace ,
\end{equation}
using typical values of the filter width $w = 123\,\mathrm{\AA}$ and
wavelength $\lambda = 11910\,\mathrm{\AA}$ (Sect.~\ref{sec:NB118_filters}).
However, since the filters are not tophats, the detection limit in line flux
will vary with wavelength across the filter.

\section{The NB118 filter curves based on laboratory measurements}
\label{sec:NB118_filters}

Searches for emission lines in the $J$ band are difficult
due to the many strong telluric emission lines
\citep[mainly due to hydroxyl,][]{Rousselot_etal:2000} at that wavelength band. Only a limited 
number of wavelength intervals are suitable \citep[e.g.][]{Nilsson_etal:2007}.
Among these is the window at 1185\,nm, which corresponds to a
Ly$\alpha$ redshift of $z = 8.8$.
This specific window is free from strong skylines within the wavelength range
from 1179\,nm to 1196\,nm and is the target of the NB118 VIRCAM filters.
These narrow-band interference filters were specified to have
a central vacuum wavelength of $1185\pm2$\,nm and
a FWHM of $10\pm2$\,nm at an operating temperature of 100\,K
and in a convergent f/3.3 beam, which was to be approximated as
a collimated beam at an angle of incidence of 7$^\circ$.
In addition, the specifications placed limits on the out-of-passband leaks:
the average transmission between  700\,nm and 1140\,nm should be below 0.1\%,
and
the average transmission between 1250\,nm and 3000\,nm should be below 0.01\%.

\begin{figure*}[htbp]
\centering
\includegraphics[width=0.90\textwidth,bb=36 20 822 422]{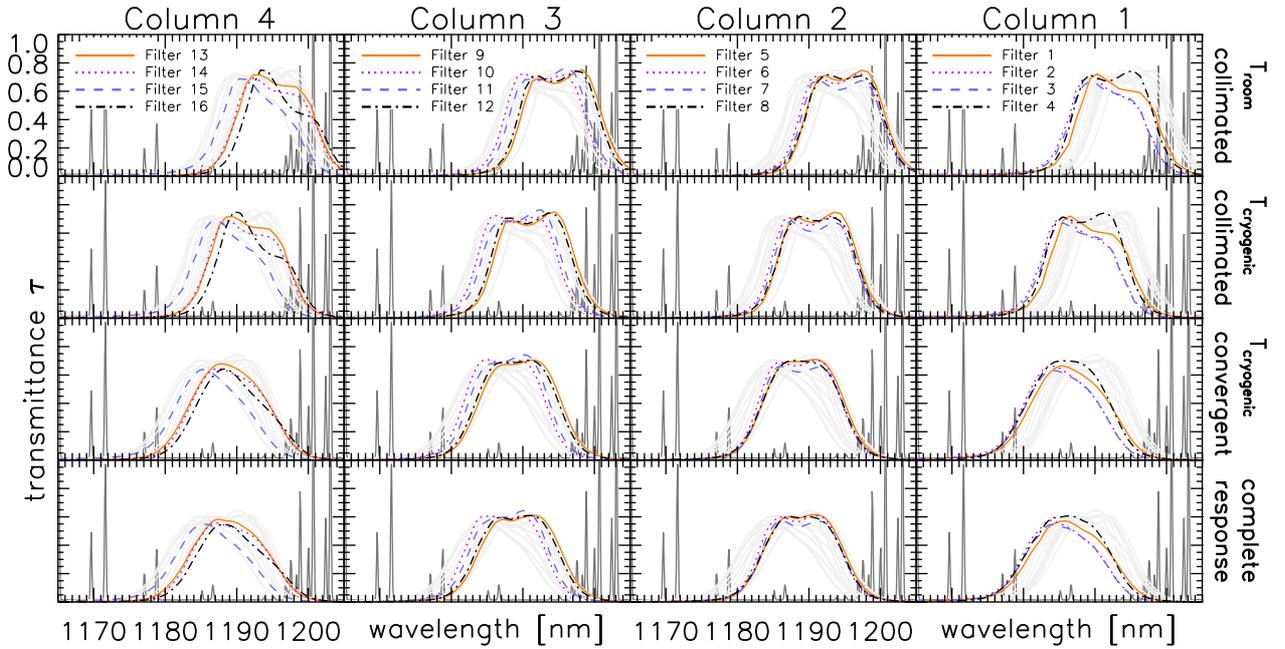}
\caption[]{%
Four different steps in the determination of transmittance curves shown for
each of the 16 individual copies of the NB118 filter in VIRCAM\@.
First row: Collimated beam measurement at room temperature as provided by the
manufacturer NDC\@.
Second row: Shift as expected due to the temperature difference.  
Third row: Theoretical conversion to the convergent beam. 
Fourth row: Response function including all theoretical filter transmittance,
mirror reflectivity, detector QE, and atmospheric transmission (PWV = 1\,mm,
airmass = 1).
In the plot, the 16 filters are separated into subsets of four filters
corresponding to the four columns or stripes of the combined image
(cf.\ Fig.~\ref{fig:pawprint}d).
In addition, the other 12 filters are shown in every panel in light gray.    
Positions and relative strengths of the sky emission lines are also included.
\label{fig:filtercurves}
}
\end{figure*}

A total of 20 individual filters 
(of size 54\,mm $\times$ 54\,mm)
were delivered in 2007 by
NDC Infrared Engineering\footnote{\url{http://www.ndcinfrared.com}},
of which 16 were installed in VIRCAM (one filter in front of each
of the 16 detectors) and four were kept as spares. With the
delivery of the filters, the manufacturer provided measurements of the
transmittance in the collimated beam at normal incidence at room
temperature over the wavelength range 1100--1300\,nm with a sampling
of 1\,nm.  The wavelength type was not specified; we have assumed it to be air.
The manufacturer stated that the central wavelength of the filters would move
down 7\,nm with cooling and cone angle, and that the bandwidth at 10\% would
increase by 0.9\,nm.

NDC has in April 2013 provided additional
information about the filters based on
recent measurements done by NDC on filter parts still in their possession,
using more accurate equipment than was available originally.
NDC now predicts that the central wavelength of the filters would move
down 5.2\,nm with cooling and cone angle.

We will now derive this shift in wavelength for the 
16 NB118 filters installed in VIRCAM\@.
When using narrow-band interference filters in fast convergent beams
under cryogenic temperature, several considerations have to be taken
into account \citep[e.g.][]{Reitmeyer:1967,Parker:1998,Morelli:1991}.
First, the passband
of the filters is temperature dependent.  In VIRCAM there is no temperature
sensor on the filters, only on the filter wheel hub. This latter temperature is
reported in the headers of the individual images, and for this dataset the
median value is 101.4\,K, with a range of 101.1--101.7\,K\@.  Paranal Science
Operations estimate that (a)~the typical filter temperature is $90\,\mathrm{K}
\pm 5\,\mathrm{K}$, and (b)~the filters typically are 5--10\,K colder than the
filter wheel hub, which for this dataset would imply a filter temperature of
91--96\,K\@.  We will assume a filter temperature of 90\,K in our calculations.
The temperature difference between room temperature (295\,K) and operating
temperature (90\,K) gives rise to a blueward shift of the passband of 3.8\,nm,
based on a linear relation of 0.0186\,nm/K measured by NDC in 2013.

Second, VIRCAM has no collimated beam and the filter wheel is located in the
fast convergent beam 
\citep[f/3.25 at the Cassegrain focus,][]{Dalton_etal:2006} of VISTA\@.
This means that effectively a
superposition of rays with various incidence angles is passing the filter.  A
ray with incidence angle away from the normal will experience a blue-shifted
passband, due to the decreased optical path difference of interfering rays
\citep[e.g.][]{Morelli:1991}.  Therefore, the effect of a convergent beam is a
further shift of the passband to shorter wavelengths compared to the normal
incidence collimated curve.  This is accompanied by a transformation of the
passband, which is mainly a broadening
\citep[e.g.][]{Lissberger:1970,Bland-Hawthorn_etal:2001}.  The shift and
transformation depends slightly on the position of the filter in the focal
plane.  Our calculation uses an effective refractive index of 2.2,
as quoted by NDC in 2013.
%
We also assume that the incidence angle of the chief-ray
can be estimated from the
position of the object as $7.15^\circ$ per degree distance from
the centre of the FOV \citep{Findlay:2012}.

Based on the first principles described above, we calculated the expected
transmittance for each of the 16 NB118 filters according to their position
in the cryogenic convergent beam of VIRCAM from the available collimated
beam measurements (more details will be given in Zabl et al., in prep.).
In the calculation we assume that both a change of temperature and a change
of incidence angle (still for a collimated beam) only shifts the
filter curve and does not change the shape.
However, measurements of filter curves in the
literature at different incidence angles show different extents of deviations
from this assumption \citep[e.g.][]{Vanzi_etal:1998,Ghinassi_etal:2002}.
Therefore, the calculated curves must be understood as an approximation.

\begin{figure*}[htbp]
\centering
\includegraphics[width=0.90\textwidth,bb=-99 281 714 503]{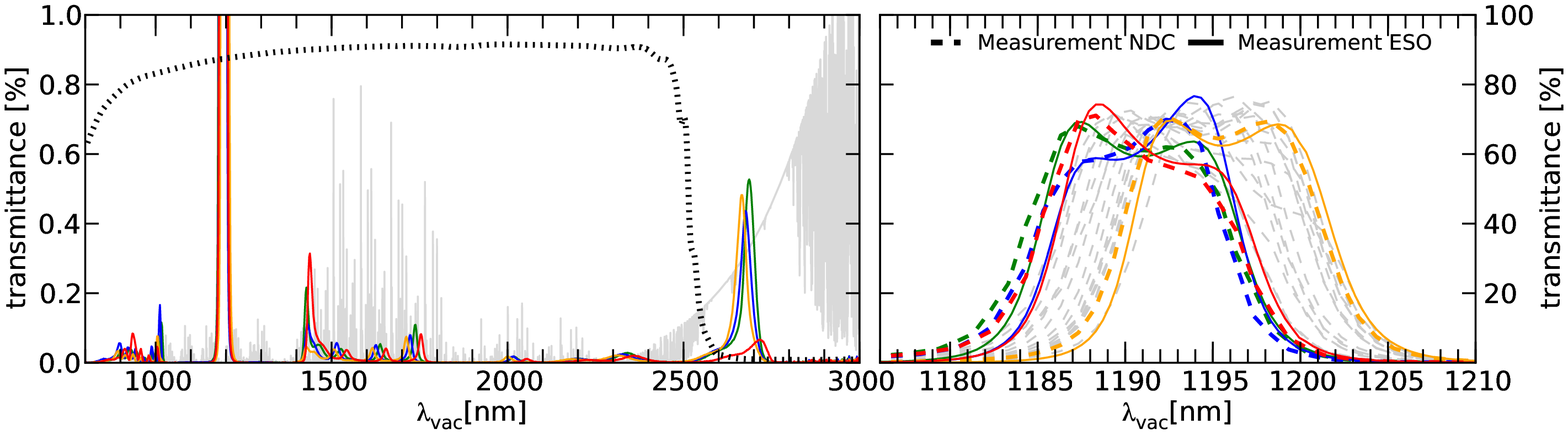}
\caption[]{%
Filter curves for the 4 spare filters.
These filters are relevant since they are the only ones that have two sets
of measurements available:
by NDC over 1100--1300\,nm (dashed lines) and
by ESO over 800--3000\,nm (solid lines).
Each colour represents one of the four filters.
The left panel shows the full wavelength range covered by our measurements
from the ESO laboratory. The grey line is a sky emission spectrum (see text),
and the dotted line represents detector QE and mirror reflectivity
(see text), but not atmospheric absorption.
The right panel shows a zoom near the passband of the filters;
here both NDC 
and ESO 
measurements are available.
The 16 grey curves in the background show the 16 filters installed
in VIRCAM (these are the filter curves shown in the top row of
Fig.~\ref{fig:filtercurves}).
All curves have been transformed from air to vacuum, but no further
transformations have been applied.
\label{fig:sparefilters}
}
\end{figure*}

In Fig.~\ref{fig:filtercurves}, all the original collimated beam measurements
performed at room temperature are shown in the first row of panels,
while the same curves shifted to 90\,K are shown in the second row.
The convergent beam transformed curves, including the temperature shift,
are shown in the third row.
Additionally, the complete response function including detector QE, primary and
secondary mirror reflectivity, and an atmosphere of airmass 1.0 and a
precipitable water vapour (PWV) of
1.0\,mm is plotted in the fourth row, using data from ESO\footnote{%
\url{http://www.eso.org/sci/facilities/paranal/instruments/vircam/inst/Filters_QE_Atm_curves.tar.gz}
--- for reference, at 1190 nm the atmospheric transmission is 99.3\%, the
detector QE is 91.6\%, and the M1 and M2 reflectivities are 97.5\% and 97.8\%,
respectively, which likely refer to the silver coating in use at the time of
the GTO observations.  The other optical elements in the system (camera
entrance window and lenses L1, L2 and L3) are not considered.}.

The total blueshift from the measurements at room temperature in a collimated
beam (Fig.~\ref{fig:filtercurves}, top row) to our calculation representing
cryogenic temperature in the convergent beam, with atmosphere, mirror
reflectivity and detector QE (Fig.~\ref{fig:filtercurves}, bottom row) is
6.0\,nm 
on average over the 16 filters (detectors), where
3.8\,nm 
comes from
temperature and 2.2\,nm 
comes from the convergent beam.  The temperature shift
is the same for all filters, whereas the shift due to the convergent beam is
larger for the filters further from the optical axis.  The total predicted
blueshift is
6.3\,nm 
for the outermost filters (1, 4, 13 and 16), and
5.6\,nm 
for the innermost filters (6, 7, 10, 11).

The 16 final cryogenic convergent beam curves, $T(\lambda)$, at their
respective positions within the beam have an average mean wavelength
$\lambda_0$
of 1187.9\,nm, with
a minimum of 1184.8\,nm (filter~2) and
a maximum of 1189.6\,nm (filter~9),
where the mean wavelength of the given filter is calculated as
\begin{equation}
	\lambda_0 = \frac{\int T(\lambda) \lambda d \lambda}
                         {\int T(\lambda) d \lambda}
\label{eq:meanwave}
\end{equation}
\citep[e.g.][]{Pascual_etal:2007}.
The FWHM of these curves is 12.3\,nm on average over the 16 filters,
and ranges from 11.5\,nm (filter~15) to 12.9\,nm (filter~4).

Due to inevitable production differences, the individual NB118 filters differ
slightly from each other. The 20 filters manufactured were carefully inspected
for obvious production problems \citep{Nilsson:2007}. One filter had a
problem and was designated as spare. The 3 filters with the most blue central
wavelengths were also designated as spare. The remaining 16 filters were
installed in VIRCAM in such a way that the 4 filters in a given column of
filters/detectors were as identical as possible. In the ``stripe'' observing
pattern used both here and in UltraVISTA for the NB118 observations,
data from different filters/detectors are only
mixed within a column (cf.\ Sect.~\ref{sec:obs} and Fig.~\ref{fig:pawprint}).
By this arrangement of the filters,
the effect of an effective bandpass broadening on the reachable line depth
and the minimum detectable equivalent width is minimised.

In this work (Sections~\ref{sec:NB118_sky_brightness} and
\ref{sec:inferred_central_wavelength}) we conclude that the 16 NB118 filters
in VIRCAM have some problems: the passband is shifted to the red,
and some filters show signs of red-leaks. For this reason the 4 spare
filters become important.
For logistical reasons it was not possible to obtain an independent
measurement of the 16 filters that were installed in VIRCAM, so for these
only the NDC measurements over 1100--1300\,nm are available.
However, we had the 4 spare filters re-measured at ESO,
allowing a check of the NDC curves. Furthermore, the ESO measurements
were performed over the wide wavelength range of 800--3000\,nm,
allowing a check of possible red-leaks in these filters.
The comparison of the NDC and ESO measurements is shown in the right panel of
Fig.~\ref{fig:sparefilters}.
Both sets of measurements agree reasonably well in shape. However,
the central wavelengths measured by NDC are about 0.7\,nm
shorter than those measured at ESO\@.
The uncertainty in the ESO measurements is estimated to be 0.4\,nm.

The full wavelength range of the ESO measurements
is shown in the left panel of Fig.~\ref{fig:sparefilters}.
In addition to some smaller leaks spread over the complete wavelength range,
substantial leaks exist near 2675\,nm
for three out of the four spare filters.
According to NDC, a leak at this wavelength
would be where the coated blocking meets the absorption on the BK7 substrate.
The average transmission in the range 1250--3000\,nm for the four
spare filters is 0.022\%, 0.019\%, 0.020\%, and 0.012\%, where the main
contribution in the first three cases is coming from the 2675\,nm
leak. A level of 0.020\% violates the specifications by a factor of two.

Although the 2675\,nm
leaks are mainly outside the efficiency range of the
detectors, thermal sky light passing through these leaks might significantly
increase the background level for these filters (had they been used in
the instrument).  We estimated the contribution from the leaks to the
sky-background based on the Gemini Observatory theoretical sky
spectrum\footnote{\url{http://www.gemini.edu/sciops/telescopes-and-sites/observing-condition-constraints/ir-background-spectra}}
calculated using \citet{Lord:1992}
for an airmass of 1.0, a PWV of 2.3\,mm and at an atmospheric
temperature of 280\,K\@.
For the calculation, we have shifted the measured transmittance curves by
6.0\,nm towards the blue to account for the expected passband shift.
The sky spectrum 
is shown in the left panel of Fig.~\ref{fig:sparefilters}.  

Then, we calculated the fraction of detected sky photons passing the filters
both in and out of the main passband.  Here, we define the filter passband
by the wavelength range 1165--1210\,nm, where by specification the
transmittance outside these interval boundaries was required to be below 1\%.
We found that if the spare filters were used in VIRCAM, sky light
passing through the out of passband leaks would contribute
38\%, 31\%, 39\%, and 26\% to the total sky background, respectively.
We further note that the wavelength range 2500--2770\,nm alone would
contribute 24\%, 21\%, 24\%, and 8\%, respectively.
These calculations should be considered as crude estimates for
several reasons.
First, the used sky spectrum seems to overpredict the inter-line telluric
background.
Second, we do not have estimates for the accuracy of most of the used data.
Third, we are using the quantum efficiency (QE) curve as used in the ESO
exposure time calculator. However, the QE is probably varying at some level
from detector to detector, which could have strong consequences on the impact
of the 2675\,nm leak (cf.\ Sect.~\ref{sec:NB118_sky_brightness_all_VISTA}).
Fourth, we are simply assuming that the leaks are shifted due to convergent
beam and temperature by the same amount as the main passband. However, a
slightly different shift of the main red-leak would give different results.

For the 16 filters installed in VIRCAM we do not have the required
filter curve measurements to predict the effect of possible red-leaks.
However, we can study these indirectly, as done in
Sect.~\ref{sec:NB118_sky_brightness}.

The above-mentioned sky spectrum has very high spectral resolution,
i.e.\ has very narrow emission lines. For aesthetic reasons we have
convolved the spectrum by a Gaussian kernel with a FWHM of 0.2\,nm
(and down-sampled from 0.02\,nm to 0.1\,nm)
to reach a spectral resolution similar to that of X-shooter
and used that for the plots (Figs.~\ref{fig:filtercurves},
\ref{fig:sparefilters} and \ref{fig:zspec_lambda}).
In Fig.~\ref{fig:sparefilters} this procedure affects 
how strong the skylines appear relative to the continuum
near 2675\,nm.

In Sect.~\ref{sec:inferred_central_wavelength} we make a first attempt
at comparing our predicted filter curves (Fig.~\ref{fig:filtercurves},
bottom row) to the narrow-band excess as function of wavelength as
inferred from spectroscopic redshifts. We find that the filter curves
need to be shifted approximately 3.5--4\,nm to the red to match the
data, while the FWHM values appear correct.
Such a shift implies that the average mean wavelength $\lambda_0$
of the 16 filters probably is about 1191--1192\,nm.
It could be argued that the filters should be renamed NB119,
but we have kept the name NB118 for consistency with the current
documentation and processing setup at ESO and CASU.

\section{The NB118 filter curves based on observations}
\label{sec:Nbexcess_results}

\subsection{Selection of candidate emission line objects}
\label{sec:selection_Nbexcess}

\begin{figure}[htbp]
\centering
\includegraphics[scale=0.65,bb=48 233 296 469]{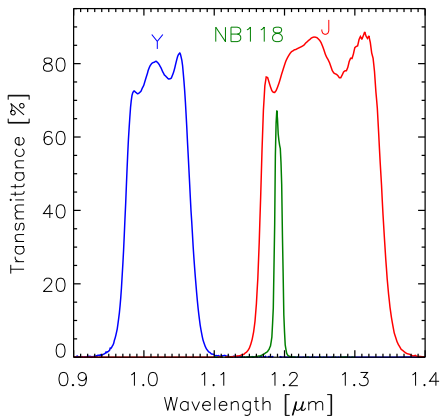}
\caption[]{%
Filter curves for $Y$, NB118 and $J$.
\label{fig:filters_Y_NB118_J}
}
\end{figure}

NB118 is located inside the J band, so the simplest selection of
candidate emission line objects would use only the $(J-\mathrm{NB118})$ colour.
NB118 is not located at the centre of the J band,
so objects with a non-flat continuum in $F_\nu$ would have a
non-zero $(J-\mathrm{NB118})$ colour just from the continuum slope.
Since NB118 is located near the blue edge of the J band,
see Fig.~\ref{fig:filters_Y_NB118_J},
it is natural to additionally use the Y-band data to select
candidate emission line objects.

\begin{figure*}[htbp]
\centering
\includegraphics[width=0.85\textwidth,bb=17 196 570 753]{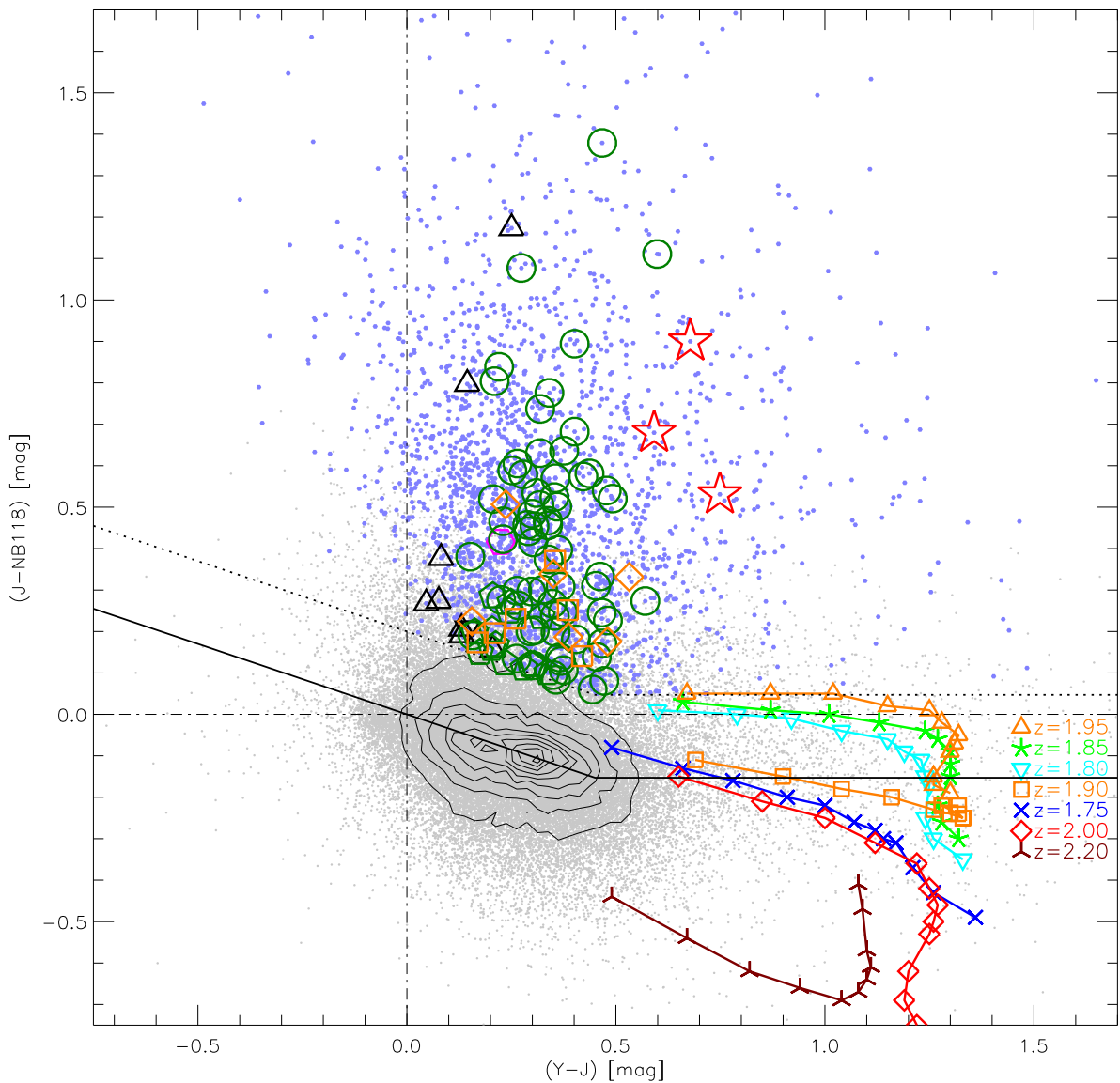}
\caption[]{%
$(J-\mathrm{NB118})$ vs $(Y-J)$.
The blue and grey dots are objects with and without significant narrow-band
excess, respectively.
The contours show the density of objects after excluding likely stars.
The blue dots additionally marked by a large symbol are line emitters
confirmed by spectroscopy, see Fig.~\ref{fig:colour_magnitude} for a legend.
The solid line is meant to represent the locus of the majority of the objects.
The dotted line located 0.2\,mag above is one of the selection criteria.
The symbols connected by coloured lines represent certain 
\citet{Maraston:2005} model galaxies (see text) without emission lines,
placed at the 7 redshifts indicated on the panel and observed at times
(from left to right)
0.2, 0.3, 0.4, 0.5, 0.6, 0.7, 0.8, 0.9, 1.0, 1.5, 2.0 and 3.0\,Gyr
(except for $z=2.2$ where 3.0\,Gyr is not shown).
\label{fig:J_NB118_vs_Y_J}
}
\end{figure*}

Figure~\ref{fig:J_NB118_vs_Y_J} shows $(J-\mathrm{NB118})$ vs $(Y-J)$
for all objects that are detected at
5$\sigma$ in NB118 and at 2$\sigma$ in $Y$ and $J$\@.
A bright cut in all 3 bands of 18\,mag has been imposed.
The objects that are candidate emission line objects according to
the criteria discussed below are shown in blue and the rest in grey.

We want to determine the typical locus of galaxies in this diagram.
We illustrate this by the contours, which show the density of objects
after excluding likely stars,
defined as objects with FWHM less than 1$''$ in the NB118 stack;
this mainly means that the contours exclude the clump of objects
(i.e.\ stars) with colours $(-0.1,0.0)$ in the figure.
For objects with $(Y-J)$ below about 0.5 the trend is clear, and we
model this with a line with slope $-0.34$, as shown by the solid line
for $(Y-J) < 0.45$.
For redder objects the situation is less clear.
To illustrate the range of $(J-\mathrm{NB118})$ colours that the continua of
galaxies can have in this part of the diagram,
we show a set of \citet{Maraston:2005} models
\cite[see also][]{Maraston:1998}
with solar metallicity, red horizontal branch morphology, Salpeter IMF
and an exponentially declining SFR with an e-folding time of 0.1\,Gyr.
These model spectra do not have emission lines.
We have redshifted the model spectra to $z$ = 1.75--2.00 as well as $z=2.20$,
as indicated on the figure.
We have used the models with ages from 0.2\,Gyr to 3\,Gyr
(from left to right in the figure; for $z=2.20$ the oldest model is 2\,Gyr).
The $z$ = 1.75--2.00 models are interesting as the $(J-\mathrm{NB118})$ colour
moves up and down with redshift, due to the location
of the 4000\,{\AA} break and several absorption lines (including Balmer,
Ca and Fe lines) in the NB118 and $J$ filters.
It is worth noting that \citet{Bruzual_Charlot:2003} models give the same
behaviour.
With the data and models in mind, we define a constant value if $-0.153$
as the ``typical'' $(J-\mathrm{NB118})$ colour for galaxies with
$(Y-J) > 0.45$,
as shown on the figure by the solid line.
The dotted line, located 0.2\,mag above the solid line, is used as one
of the criteria defining candidate emission line objects (see below).

It is noteworthy that the $z=2.2$ models in Fig.~\ref{fig:J_NB118_vs_Y_J}
are located so much below the $z=1.95$ models. This means that selecting
[\ion{O}{ii}] emitters at $z=2.2$ (where [\ion{O}{ii}] is in the NB118 filter)
is difficult, since either one sets the threshold in $(J-\mathrm{NB118})$ low
at the price of contamination by $z \approx$ 1.75--2 galaxies,
or one sets the threshold high
at the price of only selecting [\ion{O}{ii}] emitter with a large
equivalent width (EW)\@.
Using more filters than $J$ and $Y$ would help; this is beyond the scope of
this paper. Note that this paper is mainly based on results for
NB118 emitters with spectroscopy, which are mostly objects at lower
redshifts.

\begin{figure*}[htbp]
\centering
\includegraphics[width=0.85\textwidth,bb=17 196 570 753]{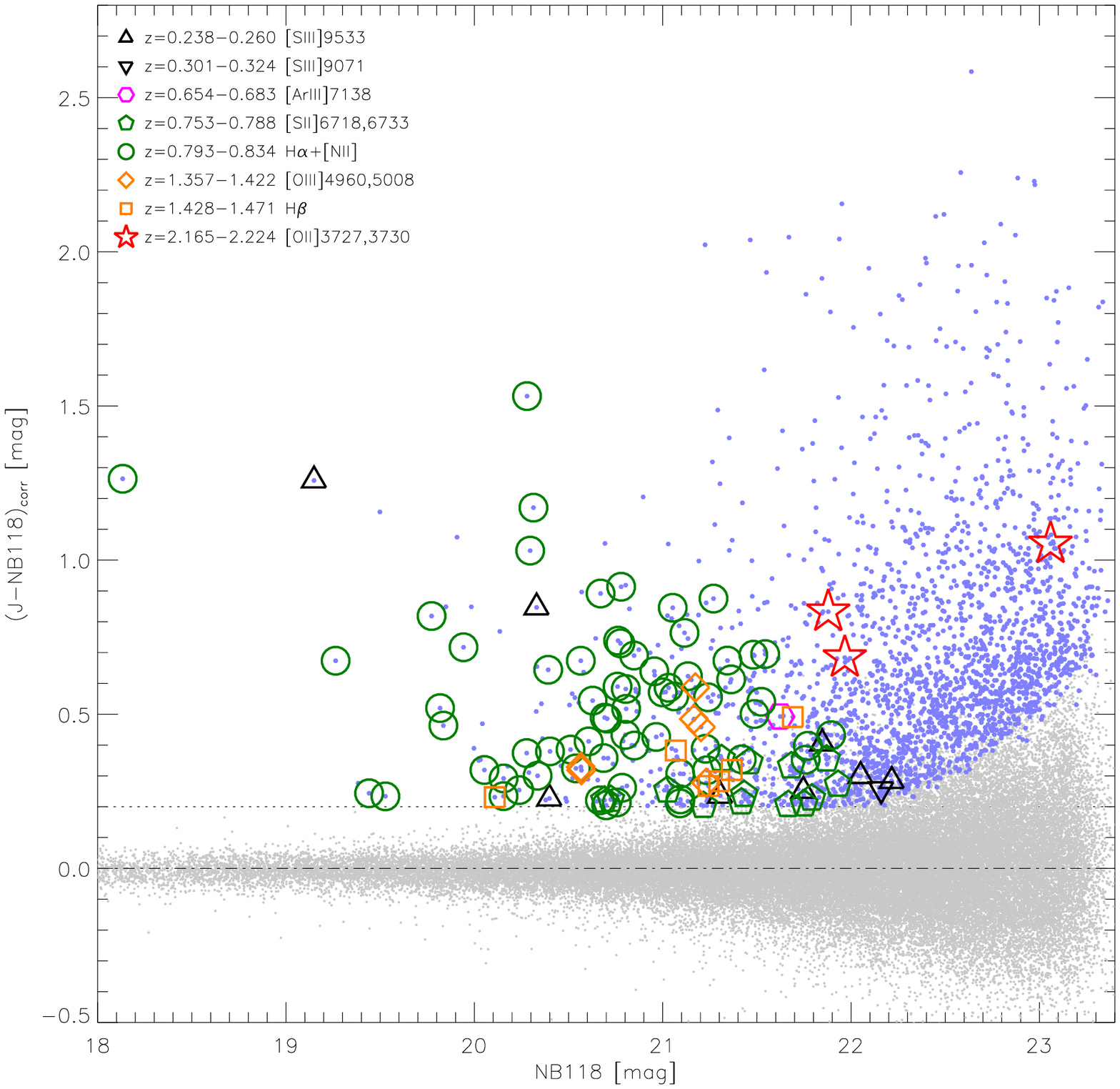}
\caption[]{%
Narrow-band colour--magnitude diagram.
The blue dots
are objects with significant narrow-band excess,
and those with a spectroscopic redshift in certain intervals are additionally
marked with large symbols, as per the legend on the figure.
The small grey dots are objects without significant narrow-band excess.
The horizontal dotted line marks one of the 
selection criteria,
see the text.
\label{fig:colour_magnitude}
}
\end{figure*}

Based on the analysis above, we define a $(J-\mathrm{NB118})$ colour
that is corrected for the fact that the NB118 filter is not located at
the centre of the J-band filter.
This quantity is denoted $(J-\mathrm{NB118})_\mathrm{corr}$ and is essentially
the $(J-\mathrm{NB118})$ colour minus the solid line in Fig.~\ref{fig:J_NB118_vs_Y_J},
specifically
%
%
\begin{equation}
(J-\mathrm{NB118})_\mathrm{corr} =
  \begin{cases}
    (J-\mathrm{NB118}) + 0.34 (Y-J) & \mbox{if\,} (Y-J) \le 0.45 \\
    (J-\mathrm{NB118}) + 0.153      & \mbox{if\,} (Y-J) >   0.45 \\
    (J-\mathrm{NB118}) + 0.07       & \mbox{if $Y$ not detected}        
  \end{cases}
\label{eq:J_NB118_corr}
\end{equation}
where the last branch refers to the object not being detected in the
$Y$-band stack at 2$\sigma$, and where the applied constant (0.07)
simply is the median correction for the rest of the objects.
(Of the 2308 candidate emission line objects defined below,
17 did not have a 2$\sigma$ $Y$-band detection.)
We note that \citet{Ly_etal:2011} used a similar functional form,
with a slope below a certain broad-band colour
--- $(z'-J) = 0.5$ in their case --- and a constant thereafter.

We can now define the criteria for selecting candidate emission line objects.
These criteria are similar to those used in many previous studies
\citep[e.g.][]{Geach_etal:2008,Shioya_etal:2008,Villar_etal:2008,
Sobral_etal:2009:Ha,Ly_etal:2011,Nakajima_etal:2012,Lee_etal:2012,
Sobral_etal:2012,Sobral_etal:2013}.

First, we require the objects to be detected at 5$\sigma$ in NB118 and
at 2$\sigma$ in $J$; we do not place any constraints on $Y$\@.
Recall that throughout this paper we use errors that are based on
the SExtractor flux errors scaled up by a factor to correct for the
effects of correlated errors introduced by the resampling
(Sect.~\ref{sec:photometry}).
In principle we could remove the requirement on $J$, but see below.

Second, we require
\begin{equation}
(J-\mathrm{NB118})_\mathrm{corr} > 0.2 \enspace .
\label{eq:selection1}
\end{equation}
This translates into a minimum observed-frame EW of 29\,{\AA}
(for filter~1 at the peak transmittance of the filter)
for the objects where the continuum colour (i.e.\ the colour in the absence
of the emission line) is well modelled by Eq.~\ref{eq:J_NB118_corr}.
This is probably the case for the line emitters at $z < 1.5$ but
not for those at $z=2.2$ (i.e.\ [\ion{O}{ii}]).

Third, we require the corrected colour to be positive
at 2.5$\sigma$ significance, i.e.\
\begin{equation}
(J-\mathrm{NB118})_\mathrm{corr} > 2.5 \,
\sigma_{(J-\mathrm{NB118})_\mathrm{corr}} \enspace ,
\label{eq:selection2}
\end{equation}
where we simply calculate the uncertainty 
$\sigma_{(J-\mathrm{NB118})_\mathrm{corr}}$
by propagating the magnitude errors on NB118 and $J$ (and $Y$ where applicable)
through Eq.~\ref{eq:J_NB118_corr}.
In the literature often a global uncertainty is used,
but we have used the individual uncertainties, which automatically
track the varying noise (depth) across the NB118 stack

The above selection produces a sample of 2308 
candidate emission line objects out of 57,882 
objects that are detected at 5$\sigma$ in NB118 and at 2$\sigma$ in $J$
(after masking of bad regions, etc.).

As stated, we have implemented the selection expressed by
Eqs.~\ref{eq:selection1} and \ref{eq:selection2}
in terms of magnitudes. We can do this since we have required
the objects to be detected in $J$\@.
In principle it is better to use the corresponding equations based on counts
rather than magnitudes \citep[e.g.][]{Bunker_etal:1995},
since then objects that are undetected in $J$
(possibly even having a slightly negative flux in $J$) are handled naturally.
In this paper, however, we mainly base our results on objects with
spectroscopy, and these are all detected in $J$\@.
For reference,
23 objects 
are detected at 5$\sigma$ in NB118 but not at 2$\sigma$ in $J$\@.
It is likely that most of these are unreal, i.e.\ due to noise or
artefacts in the NB118 image.

The selection of candidate emission line objects
has already been illustrated in the colour-colour plot in
Fig.~\ref{fig:J_NB118_vs_Y_J}.
Another view is provided by the colour-magnitude plot in
Fig.~\ref{fig:colour_magnitude}.
Note again that since we use the individual uncertainties to evaluate
the significance of the narrow-band excess (Eq.~\ref{eq:selection2}),
the used 2.5$\sigma$ selection cannot be represented by a single curve
in the colour-magnitude plot.

As an alternative to the above selection method, one could use the 
(broad1 $-$ narrow) vs (narrow $-$ broad2) colour/colour selection technique
\citep[e.g.][]{Moller_Warren:1993,Fynbo_etal:2003b,Nakajima_etal:2012}.

\subsection{Cross correlation with spectroscopic redshift catalogues}
\label{sec:zspec}

As detailed in Sect.~\ref{sec:NB118_filters} the filters were designed to
obtain the optimal bandpass using formulas to compute the effect of
cooling them to cryogenic temperatures and placing them in
the converging beam. The measured bandpasses reported in
Sect.~\ref{sec:NB118_filters}
were measured at room temperature and in a collimated beam.
To verify how well our theoretical formulae worked,
the true bandpass should be measured ``in situ'',
i.e.\ with the filters located in the instrument at
cryogenic temperature and vacuum, and with the light following the same
path as during the observations, i.e.\ meeting the same optical elements and
having the same superposition of rays from different angles.
One way to achieve this would be using a tunable monochromatic light source
combined with a device to uniformly illuminate the dome flat screen;
we do not have access to such equipment.
Another is to use emission line galaxies (and AGN) with spectroscopic
redshifts.
One of the key points in our decision to observe
in the COSMOS field was exactly that many spectroscopic
redshifts would become available in this field.

\begin{figure*}[htbp]
\centering
\includegraphics[width=0.90\textwidth,bb=25 192 570 749]{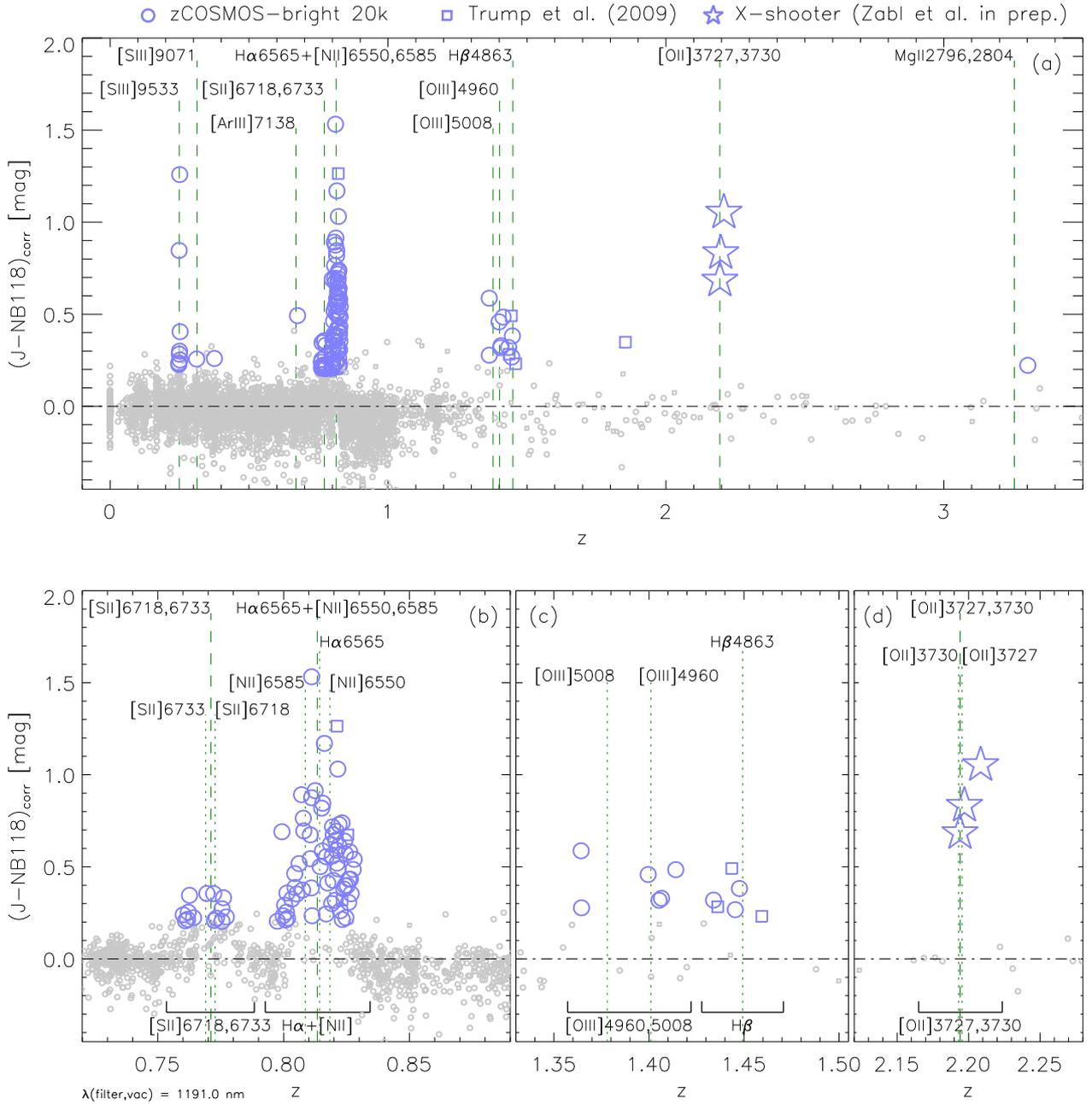}
\caption[]{%
Narrow-band colour versus spectroscopic redshift.
The large blue symbols are objects with significant narrow-band excess
(Sect.~\ref{sec:selection_Nbexcess}).
The symbol type indicates the spectroscopic catalogue, as per the legend given
on the figure.
The redshifts of strong emission lines or emission line blends selected
by the NB118 filters are marked for a typical filter vacuum wavelength of
1191\,nm (but note that the central wavelengths of the filters
in the 4 columns of the stack differ somewhat,
see the bottom row of Fig.~\ref{fig:filtercurves}).
Panels~(b), (c) and (d) provide zooms,
in which the dashed lines mark the line
blends as shown in panel~(a), while the dotted lines mark the individual lines
that are part of the line blends.
The horizontal lines mark the redshift ranges used to define the
different lines or line blends in
Table~\ref{tab:identified_line_emitters} and in
Fig.~\ref{fig:colour_magnitude}.
The low values of the colour at $z \approx 0.9$ (see panels a and b)
are due to the H$\alpha$ line being in the $J$ band but not in the NB118.
\label{fig:zspec}
}
\end{figure*}

The largest set of redshifts used here comes from the
zCOSMOS project \citep{Lilly_etal:2007}.
We use the zCOSMOS-bright 20k catalogue, which essentially is a superset of
the zCOSMOS-bright 10k catalogue \citep{Lilly_etal:2009}\footnote{%
See also
\url{http://archive.eso.org/cms/eso-data/data-packages/zcosmos-data-release-dr2.html}}.
This is based on a magnitude limited ($I_{AB} < 22.5$) sample.
We mainly use redshifts that are classified as either secure or very secure,
namely those with any of the following zCOSMOS confidence classes:
3 and 4 (stars and galaxies),
13 and 14 (broad-line AGN),
23 and 24 (serendipitous stars and galaxies), and
213 and 214 (serendipitous broad-line AGN).
In addition we use redshifts with confidence class 9, corresponding to
a single-line redshift where the spectrum does not show signs of the
object being a broad-line AGN\@.
Specifically we use the subclasses 9.5 + 9.4 + 9.3
\citep[see][]{Lilly_etal:2009}.
A detail to note on these single
line redshifts is that there are only two possible interpretations
of the line seen in the zCOSMOS optical spectrum, namely
[\ion{O}{ii}] or H$\alpha$, and therefore only two possible redshifts.
Since the two
possible solutions are vastly different they will not cause us any
problems. Either the redshift is correct and we record a line, or
the redshift is wrong and the object does not matter for our purposes.
And in fact, for the sample of candidate emission line objects with redshifts
(Table~\ref{tab:identified_line_emitters}), including
class 9 does not add any ``unidentified'' objects while adding
2 [\ion{S}{ii}] emitters,
7 H$\alpha$+[\ion{N}{ii}] emitters,
1 [\ion{O}{iii}] emitter and
2 H$\beta$ emitters.
Finally, we also use class 18 (single-line redshift, broad-line AGN):
this adds 
2 [\ion{O}{iii}] emitters and
1 H$\beta$ emitter.
The reported typical 1~sigma redshift uncertainty in zCOSMOS-bright is
$110\,\mathrm{km}\,\mathrm{s}^{-1}$.

We also use the redshift catalogue from 
the COSMOS AGN spectroscopic survey,
as published in \citet{Trump_etal:2009}.
This is based on optical spectroscopy using Magellan/IMACS and MMT/Hectospec
of X--ray selected objects. Again we only use redshifts that are
classified as secure, which corresponds to their listed confidence
classes 3 and 4.

Finally we use redshifts from early spectroscopic follow-up of
objects identified as line emitters in this work
(cf.\ Sect.~\ref{sec:selection_Nbexcess}) and
as likely to have [\ion{O}{ii}] in the NB118 filter
based on photometric redshifts from \citet{Ilbert_etal:2009}.
All the 3 objects that we observed (PI: Zabl, ESO ID 089.B-0710)
with VLT/X-shooter \citep{Vernet_etal:2011} were indeed found to be
[\ion{O}{ii}] emitters at $z=2.2$ (Zabl et al.\ in prep.).

For all redshift catalogues we performed the
cross correlation with our photometric catalogue using
the tskymatch2 routine in
STILTS\footnote{\url{http://www.starlink.ac.uk/stilts/}}
\citep{Taylor:2006}.
We used a maximum distance of 0.5$''$.
For objects with a selected redshift both from zCOSMOS and
\citet{Trump_etal:2009} we used the zCOSMOS value.
In the vast majority of these cases the redshifts are concordant.

In Fig.~\ref{fig:zspec} we plot $(J-\mathrm{NB118})_\mathrm{corr}$
versus redshifts from the 3 catalogues:
circles: zCOSMOS-bright 20k,
squares: \citet{Trump_etal:2009},
stars: X-shooter (Zabl et al.\ in prep.).
Large, blue symbols are objects with significant narrow-band excess
as defined in Sect.~\ref{sec:selection_Nbexcess},
and small, grey symbols are the remaining objects.
Panel~(a) shows the full redshift range up to $z = 3.5$,
while panels~(b) and (c) show zooms
near the redshifts of certain emission lines.
The redshifts of a number of emission lines are marked by dashed lines.
In panel~(a) certain neighbouring emission lines have been shown by a single
dashed line for clarity, e.g.\ [\ion{S}{ii}]6718,6733,
whereas panels~(b) and (c) additionally show the individual emission lines
as dotted lines, e.g. [\ion{S}{ii}]6718 and [\ion{S}{ii}]6733.
All rest-frame (vacuum) wavelengths were taken from
the Atomic Line List\footnote{\url{http://www.pa.uky.edu/~peter/atomic/}}.
The dashed and dotted lines correspond to a filter wavelength of 1191.5\,nm,
a number adjusted by eye to get a good match to the data in this figure.
In this figure data from all the 16 NB118 filters are mixed (and not every
filter has the same number of redshifts), and it should
be kept in mind that the different filters are expected to have somewhat
different wavelengths, as shown by the 16 filter curves in the bottom row of
Fig.~\ref{fig:filtercurves}.
We expand on this topic in Sect.~\ref{sec:inferred_central_wavelength}.

The main conclusion from Fig.~\ref{fig:zspec} is that the set of 16
NB118 filters do indeed function as intended,
in that objects with significant NB118 excess are almost always
located at the redshifts corresponding to strong emission lines.
The breakdown by emission line is given in
Table~\ref{tab:identified_line_emitters}.
Most of the NB118 excess objects with redshifts in the catalogues used here
are $z \approx 0.8$ H$\alpha$ emitters.

The redshift limits $z_1,z_2$ used to define the identification
with a given emission line(s) in Table~\ref{tab:identified_line_emitters}
were calculated as the range that is within
$\pm0.85 \times \mathrm{FWHM(filter)}$ from the emission line(s) in question,
using a typical filter FWHM of 12.3\,nm and a typical filter wavelength
of 1191\,nm. In reality these values vary a bit from filter to filter.
The factor of 0.85 is somewhat arbitrary, but it gives a reasonable
match to the data,
see the limits indicated on the zoom panels of Fig.~\ref{fig:zspec},
particularly panel~(b) that has many galaxies.
With this definition, 3 objects are unidentified. Two of these,
at $z=0.375$ and $z=1.854$, are
probably objects that are classified as emission line objects due to
fluctuations in the photometry.
The third object is at $z=3.302$ (Fig.~\ref{fig:zspec}a),
which is just outside
the redshift interval defining \ion{Mg}{ii}2796,2804. This object
is listed in the zCOSMOS catalogue as a broad-line AGN, so it is likely
that the object has a broad \ion{Mg}{ii} emission line that could
cause the observed NB118 excess.

\begin{table}
\caption{Spectroscopically Identified Line Emitters
\label{tab:identified_line_emitters}}
\begin{center}
\begin{tabular}{llrrr}
\hline
Emission line(s)          & Air names\tablefootmark{a}& $z_1$   & $z_2$  & $N$ \\ \hline
{[\ion{S}{iii}]9533     } & {[\ion{S}{iii}]9531     } & 0.238   & 0.260  &  8 \\
{[\ion{S}{iii}]9071     } & {[\ion{S}{iii}]9069     } & 0.301   & 0.324  &  1 \\
{[\ion{Ar}{iii}]7138    } & {[\ion{Ar}{iii}]7135    } & 0.654   & 0.683  &  1 \\
{[\ion{S}{ii}]6718,6733 } & {[\ion{S}{ii}]6716,6731 } & 0.753   & 0.788  & 14 \\
{H$\alpha$+[\ion{N}{ii}]} & {H$\alpha$+[\ion{N}{ii}]} & 0.793   & 0.834  & 66 \\
{[\ion{O}{iii}]4960,5008} & {[\ion{O}{iii}]4959,5007} & 1.357   & 1.422  &  6 \\
{H$\beta$               } & {H$\beta$               } & 1.428   & 1.471  &  6 \\
{[\ion{O}{ii}]3727,3730 } & {[\ion{O}{ii}]3726,3729 } & 2.165   & 2.224  &  3 \\
{\ion{Mg}{ii}2796,2804  } & {\ion{Mg}{ii}2796,2803  } & 3.211   & 3.297  &  0 \\
{Unidentified           } & {Unidentified           } & \ldots  & \ldots &  3 \\ \hline
\end{tabular}
\tablefoot{%
The table contains the candidate emission line objects
with secure spectroscopic redshifts (see text).
Each row corresponds to an emission line or a group of emission lines
defined by the redshift interval $z_1,z_2$.
The last row gives the number of objects that do not fall into any of
these redshift intervals; one of these objects is likely an
\ion{Mg}{ii} emitting AGN, see text.
\tablefoottext{a}{Names using the air wavelengths of the emission lines}
}
\end{center}
\end{table}

\subsection{Inferred central wavelength of the filters}
\label{sec:inferred_central_wavelength}

The plot of $(J-\mathrm{NB118})_\mathrm{corr}$ versus redshift
(Fig.~\ref{fig:zspec})
suggests that we may
infer the central
wavelengths of the 16 NB118 filters from emission-line objects
with spectroscopic redshifts.
The calculation of observed-frame wavelength from redshift
($\lambda_\mathrm{obs} = (1+z) \lambda_\mathrm{rest}$)
is most simply done for strong single (isolated) emission lines
or for emission line blends where the constituent
emission line are close in redshift.

\begin{figure*}[htbp]
\centering
\includegraphics[width=0.90\textwidth,bb=6 193 570 751]{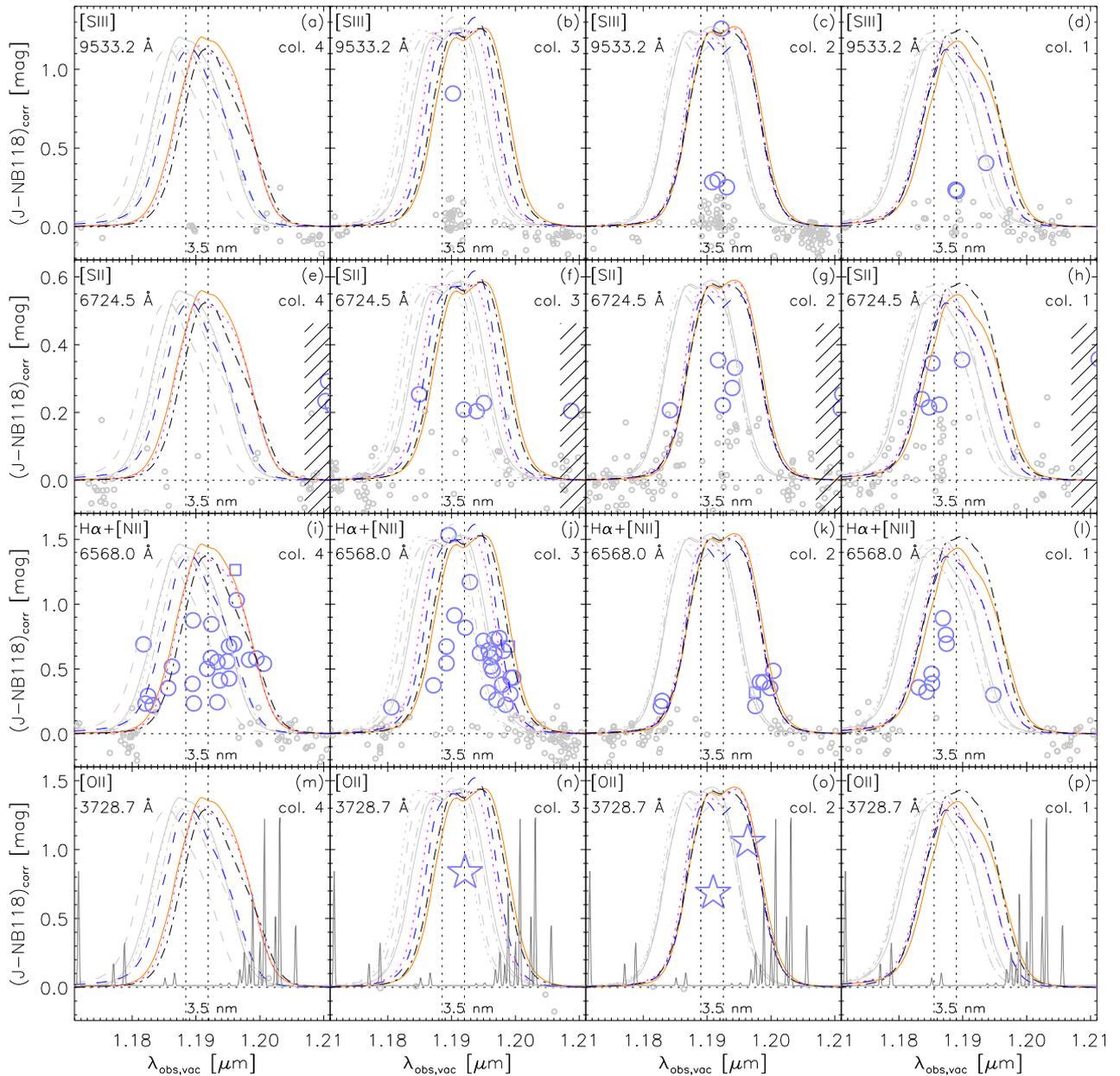}
\caption[]{%
Wavelength response of the 16 NB118 filters, as illustrated by objects
with spectroscopic redshifts. These plots are as Fig.~\ref{fig:zspec},
except that 
$\lambda_\mathrm{obs,vac} = (1+z) \lambda_\mathrm{rest,vac}$
rather than $z$ is on the $x$-axis.
The 4 rows of the figure correspond to 4 emission lines or emission line blends,
with the used rest-frame vacuum wavelength listed.
The 4 columns of the figure correspond to the 4 columns of the stack
(see Fig.~\ref{fig:pawprint}), where each column contains data from
4 NB118 filters (namely filters 1--4, 5--8, 9--12 and 13--16 for columns
1, 2, 3 and 4, respectively), and where the corresponding filter curves
are overplotted. The grey filter curves are those from the bottom row of
Fig.~\ref{fig:filtercurves}, while the coloured ones are the curves
shifted by 3.5\,nm to the red to visually match the data.
The vertical dotted lines indicate the centre wavelength of the 4 filters
in the given panel before and after the shift.
In the [\ion{S}{ii}] panels (panels~e--h) the hatched region indicate
a redshift range where narrow-band excess is likely due to H$\alpha$ rather
than [\ion{S}{ii}].
In the bottom row the theoretical sky spectrum used in
Sect.~\ref{sec:NB118_filters} is shown, scaled to fit the panel.
\label{fig:zspec_lambda}
}
\end{figure*}

For our analysis we will use the single line [\ion{S}{iii}]9533 and
three line blends, for which the effective wavelengths are derived as follows.
For the [\ion{S}{ii}]6718,6733 and [\ion{O}{ii}]3727,3730 doublets, where the
intensity ratio depends on the electron density $n_e$, we have used
[\ion{S}{ii}]6718/[\ion{S}{ii}]6733 = 1.3 and
[\ion{O}{ii}]3730/[\ion{O}{ii}]3727 = 1.3, which in both cases correspond to
$n_e \approx 10^2\,\mathrm{cm}^{-3}$, which is a typical value
\citep{Osterbrock_Ferland:2006}.
The resulting effective wavelength 
for [\ion{S}{ii}] is 6724.5\,{\AA} and
for [\ion{O}{ii}] is 3728.7\,{\AA}.
For the H$\alpha$+[\ion{N}{ii}] blend of 3 lines,
the doublet ratio [\ion{N}{ii}]6583/[\ion{N}{ii}]6548 is 
essentially fixed by atomic physics to a value close to 3
\citep[e.g.][]{Storey_Zeippen:2000},
whereas the [\ion{N}{ii}]6583/H$\alpha$ ratio depends on the metallicity and
on the source of photoionization and varies substantially.
We have used [\ion{N}{ii}]6583/H$\alpha = 0.3$,
which in terms of
$N2 \equiv \log \left( [\ion{N}{ii}]6583/\mathrm{H}\alpha \right)$
\citep[e.g.][]{Denicolo_etal:2002}
corresponds to $N2 \approx -0.5$, which is a reasonable mean value
\citep[e.g.][]{Kennicutt:1992,Gallego_etal:1997,Kauffmann_etal:2003:AGN,
James_etal:2005,Moustakas_Kennicutt:2006b,Pascual_etal:2007,
Perez-Montero_etal:2013,Hopkins_etal:2013}.
The resulting effective wavelength for H$\alpha$+[\ion{N}{ii}] is 6568.0\,{\AA}.

Ideally we would like to analyse the data for each detector
(i.e.\ each NB118 filter) separately, but in our combined image the data from
different detectors are to some extent mixed within the 4 columns (stripes)
of the image, and the number of objects with redshifts is not really large
enough to allow such a split.
We therefore only split the data into the 4 columns. It should also be noted
that the 4 filters in a given column were chosen to be as identical as possible.

The plots of $(J-\mathrm{NB118})_\mathrm{corr}$ versus
inferred observed-frame wavelength is given in Fig.~\ref{fig:zspec_lambda}.
The 4 rows of the figure correspond to the 4 emission lines/blends used
([\ion{S}{iii}]9533, [\ion{S}{ii}], H$\alpha$+[\ion{N}{ii}] and [\ion{O}{ii}]),
which therefore also correspond to 4 redshift slices.
The 4 columns of the figure correspond to the 4 columns of the combined image.
The underlying data and the symbols in Fig.~\ref{fig:zspec_lambda}
are the same as in Fig.~\ref{fig:zspec}.
The grey filter curves (identical to the coloured filter curves
in the bottom row of Fig.~\ref{fig:filtercurves})
represent our calculation of where the filters should be, as operated
in the instrument, i.e.\ at cryogenic temperature and in the converging beam.
It is seen that these curves do not provide a good match to the data:
in many panels there are objects with significant narrow-band excess
on the red side of the filters and a lack of such objects on the blue side.
If these filter curves are shifted by 3.5\,nm to the red, as shown by
the coloured filter curves, a much better match to the data is provided.

The 3.5\,nm shift was found to be the best overall value that visually
matches the data in Fig.~\ref{fig:filtercurves},
with an uncertainty that probably is 1 or 2 nm.
We have also, for each of the 4 columns, calculated the weighted mean
wavelength of the galaxies, using $(J-\mathrm{NB118})_\mathrm{corr}$ as weight,
and calculated the difference with respect to the mean predicted filter
wavelengths in that column. The resulting shifts to the red for
columns 1, 2, 3 and 4 were
$1.7 \pm 0.7\,\mathrm{nm}$,
$4.8 \pm 1.0\,\mathrm{nm}$,
$5.3 \pm 0.7\,\mathrm{nm}$ and
$4.3 \pm 1.1\,\mathrm{nm}$, respectively,
with the errors calculated using the bootstrap method.
Over all columns the shift is $4.0 \pm 0.4\,\mathrm{nm}$.

We therefore conclude that there is a difference between the 
predicted filter curves and the actual achieved filter response on the sky. 
The data are compatible with the shift being the same for all 4 columns
and for all 16 filters, but filter-to-filter differences could exist.
We have re-checked all material related to the filter definitions but
we have not been able to determine the cause of this shift.
As detailed in Sect.~\ref{sec:NB118_filters}, the calculated
total blueshift from the measurements at room temperature in a collimated beam
to the operating temperature in the convergent beam is 6.0\,nm
on average over
the 16 filters. The observed 3.5--4\,nm shift to the red could either imply
that the blueshift due to temperature and convergent beam is only
2.0--2.5\,nm instead of 6.0\,nm, which is difficult to understand,
or that there is some other effect at play, either some properties of
the VISTA/VIRCAM optics that we have overlooked, or that the filters
have been affected by their environment somehow (e.g.\ by the way they
are mounted or by the long exposure to vacuum).
We note the case of the SDSS filters that changed wavelength,
see \citet[][their Sect.~4.5]{Stoughton_etal:2002} and \cite{Bessell:2005}.
These filters have a red edge defined by
a multilayer interference edge coating (acting as a short-pass filter).
It is believed that dehydration due to the filters being placed
in vacuum caused the effective refractive index to decrease,
causing a blueshift.
It is not clear that this is relevant for the VIRCAM NB118 filters
(where we observe a shift to the red), but it does indicate that
the properties of filters can be affected by their environment.

As far as the widths of the filters are concerned, the achieved
widths indicated by the data in Fig.~\ref{fig:zspec_lambda}
are in reasonable agreement with the predictions.

A limitation of the above analysis is that there is a degeneracy between
intrinsic EW of the object and filter transmission, i.e.\ an object
with a low value of $(J-\mathrm{NB118})_\mathrm{corr}$ could either indicate
an object with an intrinsically low EW or an object with an emission line
that is at a wavelength where the filter has a low transmission.
However, averaged over many objects, such effects will diminish.
Also cosmic variance is an issue, although there should be little correlation
between the 4 used redshift slices.
A much improved analysis could be performed if we had objects with
a spectroscopically measured EW of the emission line(s) in the NB118 window.
With such data each filter curve could be derived with much better accuracy.

\section{NB118 sky brightness based on observations}
\label{sec:NB118_sky_brightness}

\subsection{NB118 sky brightness in the GTO data}
\label{sec:NB118_sky_brightness_GTO}

\begin{figure*}[htbp]
\centering
\includegraphics[width=0.90\textwidth,bb=7 187 541 750]{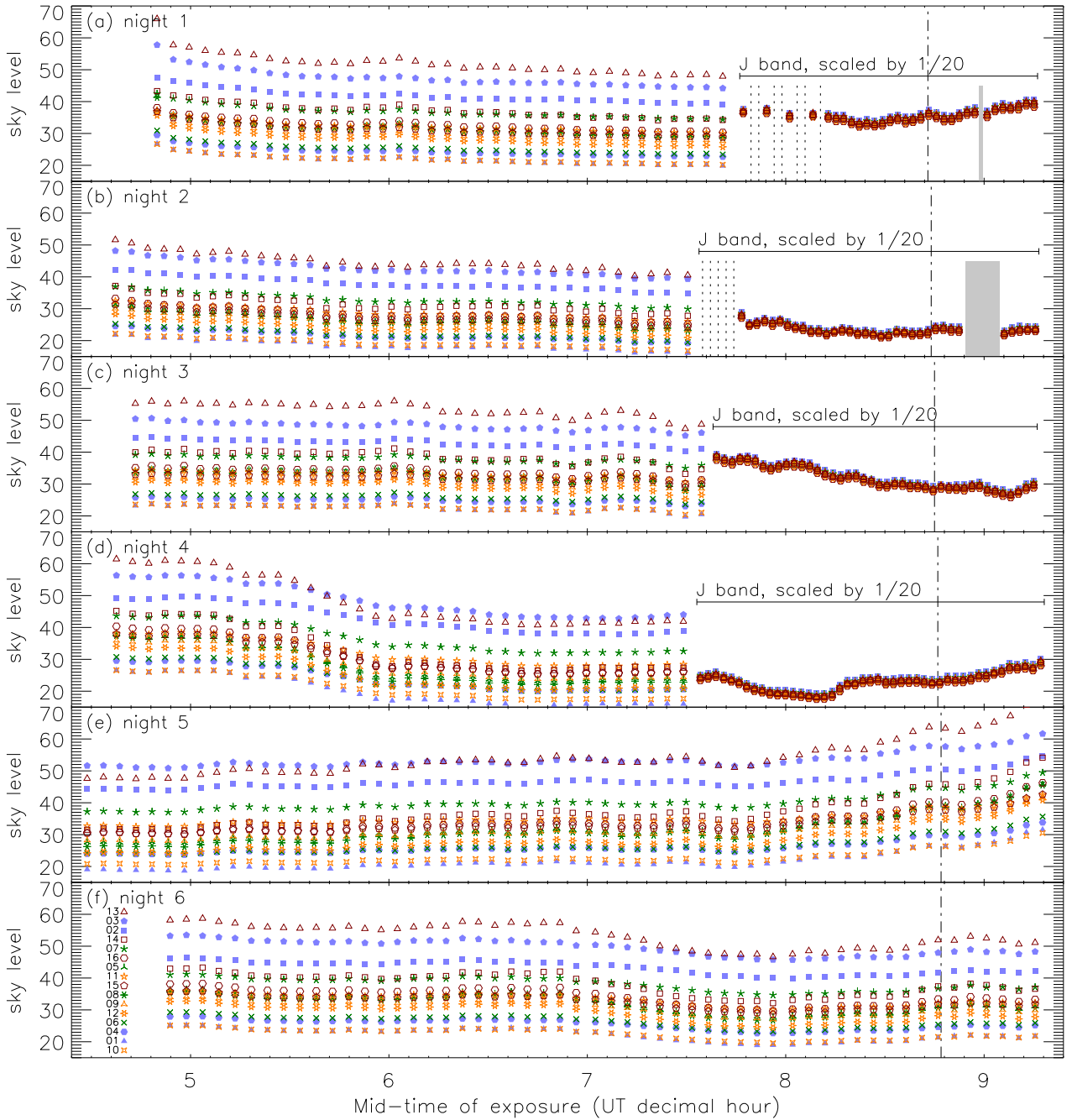}
\caption[]{%
Sky level in $e^-\,\mathrm{s}^{-1}\,\mathrm{px}^{-1}$
(with $1\,\mathrm{px}$ being about $0.34''$ on the side)
as function of time for the 6~half-nights of the GTO observing run.
The data were taken in either NB118 or $J$ band; the $J$-band sky level values
have been scaled down by a factor of 20 to fit on the plot, as indicated
on the panels.
For each exposure, i.e.\ for each point in time,
the sky level in the 16 detectors is plotted;
a symbol legend is provided in panel~(f).
In VIRCAM each detector has its own filter, and filter-to-filter variations
within the set of the 16 NB118 filters are likely the reason for the large
detector-to-detector variations seen for the NB118 sky levels.
Dot-dashed lines mark the start of twilight. 
In panels (a) and (b),
solid grey regions mark episodes where the OBs had to be restarted
due to minor technical (software) problems,
and dotted lines mark obtained $J$-band exposures for which CASU did not
provide reduced images (see Sect.~\ref{sec:CASU}).
The UT date for the observations in panel (a) is 2010--01--18, and each
subsequent panel is one day later.
%
\label{fig:skylevel_vs_time}
}
\end{figure*}

As detailed in Sect.~\ref{sec:NB118_filters} the VIRCAM
design required us to produce 16 separate narrow-band filters,
namely one for each
detector. Our design goal was to obtain 16 identical filters and as
shown (Fig.~\ref{fig:filtercurves})
the pre-installation measurements showed variations in bandpass which
were of a magnitude that should not cause any issues. We
therefore did not expect large variations in the sky brightness level
from one detector to the next. In Fig.~\ref{fig:skylevel_vs_time}
we show the actual sky level in $e^-\,\mathrm{s}^{-1}\,\mathrm{px}^{-1}$
for all the individual GTO images plotted as a function of time.
Each detector is plotted with a separate symbol/colour combination.
The $J$-band values have been scaled down by a factor of 20 to fit on the
plot; this factor reflects both the much larger width and the somewhat
brighter sky of the $J$ band compared to the NB118.
The used sky levels are those listed in the headers, computed by CASU using 
an iterative k-sigma clipped median including masking of bad pixels but
not objects, with the iteration taking care of most of the object pixels
(M. Irwin, priv.\ comm.\ 2012). The values were transformed from
gain-normalised ADU to electrons assuming a common gain of
$4.19\,e^-\,\mathrm{ADU}^{-1}$, cf.\ Sect.~\ref{sec:CASU}
and the CASU web site\footnote{\url{http://casu.ast.cam.ac.uk/surveys-projects/vista/technical/vista-gain/vista-gains}}.
This is only approximately correct, since there are detector-to-detector
differences in overall QE.

Fig.~\ref{fig:skylevel_vs_time} shows that the narrow-band background level
varies substantially from one detector to the next.
The extremes are represented by detectors 1 and 13,
with detector 13 on average having a 2.43 times higher level than detector 1,
see Table~\ref{tab:sky_level}.
The relative background level between any two detectors remains fairly
constant in time (but is not fully constant, see below).
By contrast, the spread in background levels in the $J$ band is much smaller,
as is evident from Fig.~\ref{fig:skylevel_vs_time}.
This raises the question what causes the large variations in the background
level of the narrow-band filter. 

\begin{table}
\caption{Relative NB118 background level in the GTO data as function of
detector number
\label{tab:sky_level}
}
\centering
\begin{tabular}{cccc}
\hline
\hline
   det 13 & det  9 & det  5 & det  1 \\
    2.43  &  1.41  &  1.46  &  1.00  \\
\hline
   det 14 & det 10 & det  6 & det  2 \\
    1.71  &  1.03  &  1.20  &  2.05  \\
\hline
   det 15 & det 11 & det  7 & det  3 \\
    1.48  &  1.55  &  1.78  &  2.32  \\
\hline
   det 16 & det 12 & det  8 & det  4 \\
    1.54  &  1.30  &  1.43  &  1.17  \\
\hline
\end{tabular}
\tablefoot{%
The detectors are listed in the way they appear on the sky (N up, E left).
The mean count rate in detector~1 is
$21.2\,e^-\,\mathrm{s}^{-1}\,\mathrm{px}^{-1}$
(referring to the native scale of $0.34''\,\mathrm{px}^{-1}$),
and the count rate for the other detectors can be found by scaling with the
numbers given in the table.
}
\end{table}

One possible explanation for the variations of the background level is
stray light that finds a way to the detector without passing through the
filter. This effect is a potentially serious problem for any narrow-band
imaging \citep[][their Sect.~2]{Fynbo_etal:1999}. If this were the cause, we
would expect the background level to have a systematic pattern over the 4 by 4
field spanned by the detectors in the pawprint. However, no obvious pattern
is seen (see Table~\ref{tab:sky_level}).
Furthermore, stray light going around the filters is also strongly excluded
by an examination of dark frames taken with the dome lamps on,
where essentially nothing is seen, meaning that the leak around the filters is
well below 1 photon in 100,000 (W. Sutherland, priv.\ comm.\ 2012).
The only remaining explanation for the enhanced sky background are
imperfections in the individual filter bandpass. This could be in the form of
a too wide or shifted bandpass such that airglow lines enter the bandpass,
or a too large red-leak. 
Below we show evidence that some of the filters have a noteworthy red-leak.
And in Sec.~\ref{sec:inferred_central_wavelength} we find
that the effective filter curves on the sky are not those predicted by the
laboratory measurements of filter curves.

\begin{figure*}[!ht]
\centering
\includegraphics[width=0.90\textwidth,bb=0 182 570 751]{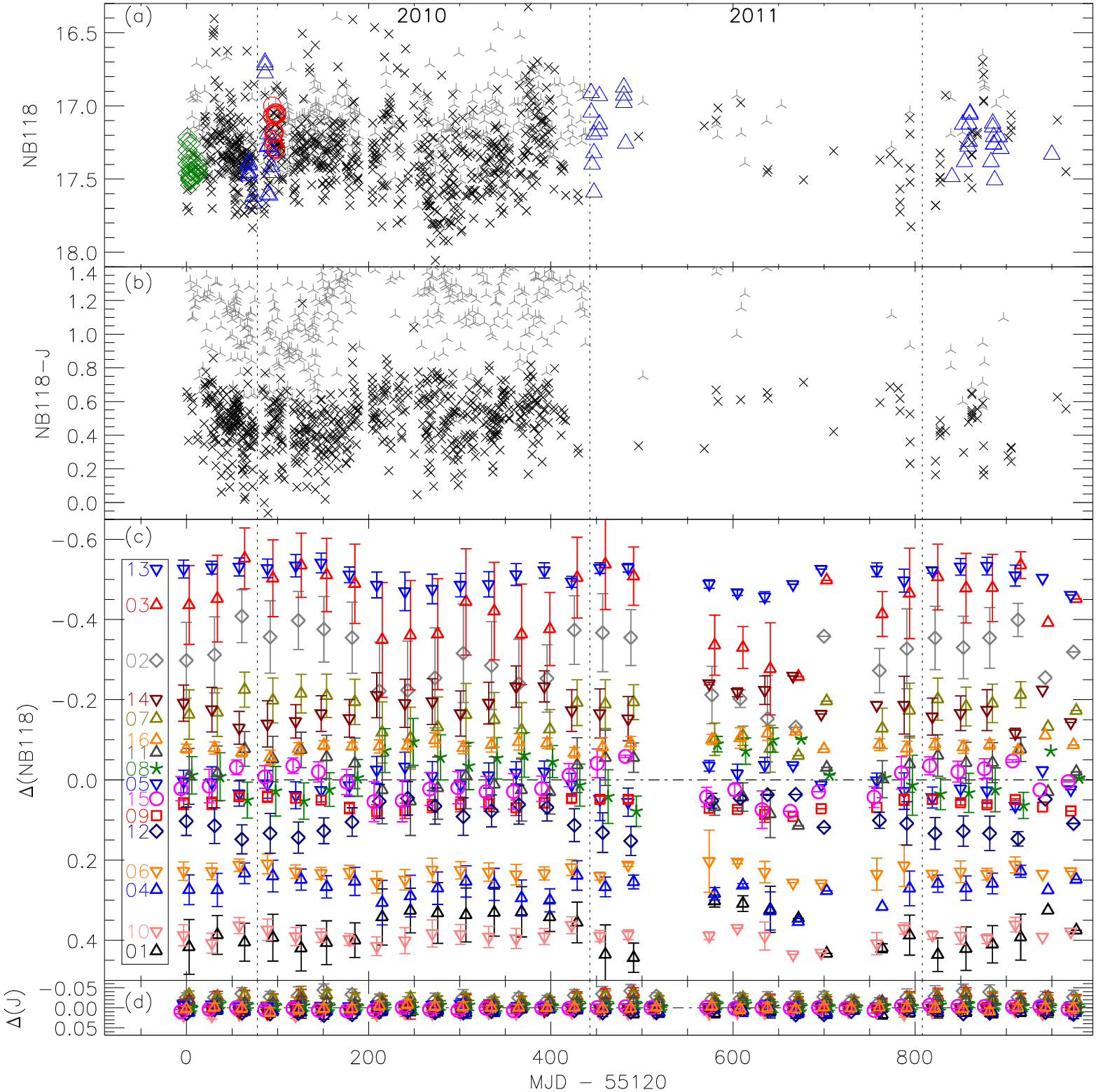}
\caption[]{%
Sky brightness in AB mag/arcsec$^2$ versus time in days
for all VISTA observations in the CASU QC tables.
Panel~(a) shows the median NB118 sky brightness,
with the median taken over the 16 detectors, i.e.\ over the 16 individual
NB118 filters.
Green diamonds: SV, blue triangles: UltraVISTA, red circles: GTO,
grey skeletal symbols: standard stars taken at the start of the night
(either in evening twilight or in the first 90 minutes of the night),
black crosses: standard stars taken the rest of the time.
Each point represents the stack made by CASU of the data coming from one OB
and one pawprint position\@.
There are therefore 18 GTO points from the 6 GTO OBs (one per night)
each observing 3 pawprint positions.
Panel~(b) shows the $(\mathrm{NB118}-J)$ sky brightness difference for
standard star observations taken in these two filters almost
simultaneously. Again the median values over the 16 detectors are used,
and symbols are as in panel~(a).
Panel~(c) shows the NB118 sky brightness in the 16 individual
detectors/filters minus the median NB118 sky brightness
(with the latter being what is plotted in panel~a).
The median value in bins of one calendar month is plotted.
The shown errorbar represents the standard deviation (if it could be computed).
A legend is provided on the panel.
The sense of the $y$-axis is such that e.g.\ detector~13 has a brighter sky
and detector~1 has a darker sky.
Panel~(d) is as panel~(c) but for $J$\@. These two panels have an
identical scale on the $y$-axis in terms of mag/arcsec$^2$ per cm on the page.
The CASU magnitudes were transformed to AB using
NB118(AB) = NB118(Vega) + 0.853 and
$J$(AB) = $J$(Vega) + 0.937.
\label{fig:skylevel_CASU_QC}
}
\end{figure*}

\subsection{NB118 sky brightness in all VISTA data and red-leak analysis}
\label{sec:NB118_sky_brightness_all_VISTA}

We have further investigated the NB118 background level using the
CASU QC tables\footnote{\url{http://casu.ast.cam.ac.uk/vistasp/qctables}},
which for all VISTA data taken since the Science Verification (SV)
in 2009 October
list a wealth of quantities, including the sky brightness in
mag/arcsec$^2$.
The images used are not the individual images (as plotted in
Fig.~\ref{fig:skylevel_vs_time}), but rather the CASU stacks of images
from a single OB and pawprint position.
The CASU sky brightness (transformed to AB using conversions calculated by
CASU\footnote{\url{http://casu.ast.cam.ac.uk/surveys-projects/vista/technical/filter-set}})
versus time
is plotted in Fig.~\ref{fig:skylevel_CASU_QC}.
Panel~(a) shows the median NB118 sky brightness, with the median being
computed over the 16 detectors.
The median values over time are as follows:
GTO data (red circles): 17.1,
SV (green diamonds): 17.4,
UltraVISTA (blue triangles): 17.3,
standard star fields, start of the night (grey skeletal symbols): 17.2,
and
standard star fields, rest of the night (black crosses): 17.3,
all in mag/arcsec$^2$.
Panel~(b) shows the $(\mathrm{NB118}-J)$ sky brightness difference
(again using the median values of the 16 detectors) for
all standard star observations where the same field was observed in the
two bands within a few minutes.
The median value over time is 1.2 and 0.5\,mag/arcsec$^2$
for observations at the start of the night and in the rest of the night,
respectively, with the division being made at 90 minutes into the night.
This means that the $J$-band sky shows a stronger change from bright to dark
from the start to rest of the night than the NB118 sky does.
Panel~(c) shows the NB118 sky brightness in the individual detectors,
with the median value subtracted. Bins of one calendar month have been used.
The pattern is fairly constant over time.
The typical min-max span is 0.95\,mag/arcsec$^2$,
in agreement with the factor of 2.43 span in the sky level for the GTO data
(Table~\ref{tab:sky_level}).
The shown errorbars indicate a robust measure of the standard deviation.
Particularly detectors 2 and 3 show a large standard deviation, meaning
that they behave less like the other detectors
(a behaviour that can also be seen e.g.\ in Fig.~\ref{fig:skylevel_vs_time}e).
Panel~(d) shows the same for the $J$ band, and here the variation is
more than an order of magnitude smaller.

The large detector-to-detector variation in the NB118 (but not in $J$-band)
sky brightness means that for the above-mentioned
standard star observations done at least 90 minutes into the night,
detector~1 has a sky brightness in NB118 of 17.8\,mag/arcsec$^2$
that is on average 0.9\,mag/arcsec$^2$ darker than that in $J$, whereas 
detector~13 has sky brightness in NB118 of 16.8\,mag/arcsec$^2$
that is essentially identical to that in $J$.

\begin{figure}[h]
\centering
\includegraphics[scale=0.85,bb=24 323 227 750]{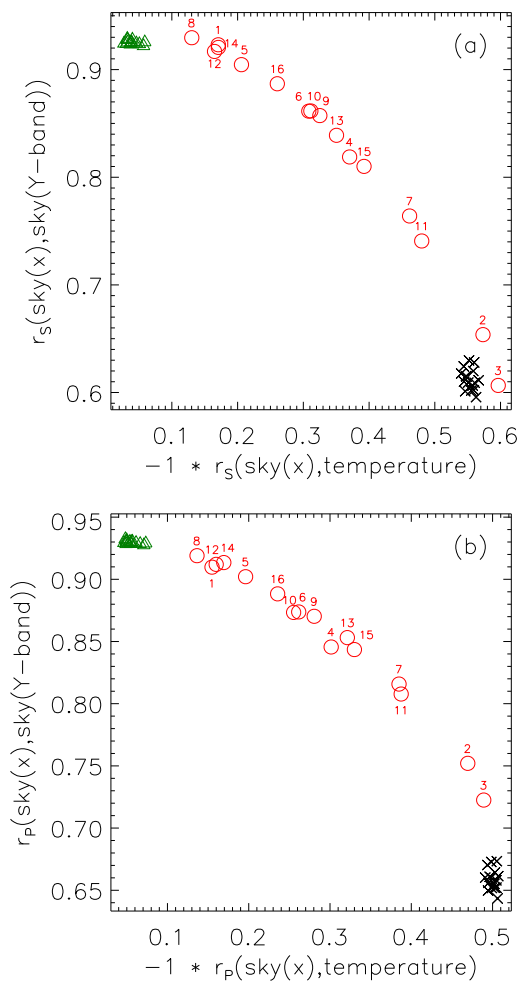}
\caption[]{%
Sky brightness correlation coefficients for the 16 detectors.
Green triangles: $J$, red circles: NB118, black crosses: $K_\mathrm{s}$.
The $x$-axes show
the correlation coefficient with the ambient temperature,
while the $y$-axis shows the correlation coefficient with the
sky brightness in the Y band in the given detector.
The correlation coefficients on the $x$-axes have been multiplied by
minus one, since they would other\-wise be negative due to the the sky
brightness being in mag/arcsec$^2$.
Panel~(a) shows the Spearman rank order correlation coefficient $r_S$
while panel~(b) shows the linear Pearson correlation coefficient $r_P$.
The individual NB118 points are labelled with the detector number.
\label{fig:sky_correlation}
}
\end{figure}

The CASU QC database contains 1080 observations
of standard stars where the same field was observed in
$H$, $K_\mathrm{s}$, $J$, $Y$, $Z$ and NB118 (in that order) within 15 minutes.
For each detector we have quantified the correlation between the
sky brightness in these 6 filters and between the ambient temperature,
by means of Spearman rank order correlation coefficients $r_S$
and linear Pearson correlation coefficients $r_P$
\citep[e.g.][]{Press_etal:1992}.
Some of these coefficients are shown in Fig.~\ref{fig:sky_correlation}.
The two panels differ only in the fact that panel~(a) shows $r_S$
while panel~(b) shows $r_P$.
The $y$-axis shows the correlation coefficient between the sky brightness
in the given band and the sky brightness in the $Y$ band, where the given band
is either $J$ (green triangles), NB118 (red circles, labelled by the detector
number) or $K_\mathrm{s}$ (black crosses).
The $x$-axis shows minus one times the correlation coefficient between
the sky brightness in the given band and the ambient temperature.
Positive values indicate that the sky gets brighter (i.e.\ smaller
in mag/arcsec$^2$) with higher temperature.
It is seen that the $J$-band sky brightness correlates strongly with the 
$Y$-band sky brightness and not strongly with ambient temperature,
whereas the opposite is the case for the $K_\mathrm{s}$-band sky brightness,
as expected.
Interestingly, the 16 NB118 filters almost span the full range between the
$J$ and the $K_\mathrm{s}$ filters. At one extreme, NB118 filters 2 and 3
essentially show as strong a correlation with ambient temperature as the
$K_\mathrm{s}$-band filters, indicating that these NB118 filters have
substantial red-leaks. These two filters were also seen to deviate from the
rest of the NB118 filters in the analysis discussed above.

\begin{figure}[h]
\centering
\includegraphics[scale=0.85,bb=24 323 227 750]{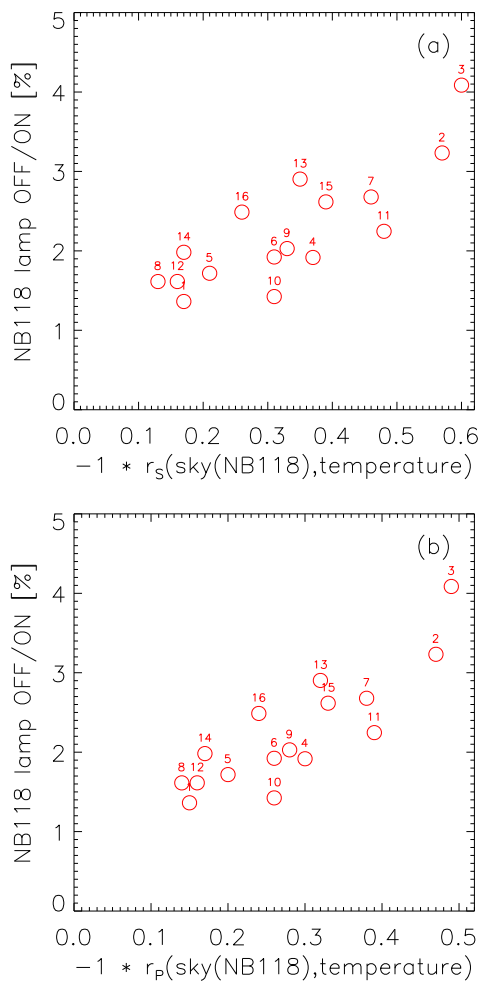}
\caption[]{%
Two measures of the red-leak for the 16 NB118 filters.
The $y$-axis shows the level in screen flats obtained with the lights
in the dome off (i.e.\ with the light source being the thermal emission
from the dome flat screen, the dome, etc.)
divided by the level in regular screen flats (i.e.\ with the dome flat screen
illuminated by a lamp). The levels have been dark subtracted.
The $x$-axes show correlation coefficients between
the NB118 sky brightness in the given detector (i.e.\ filter)
and the ambient temperature using all the VISTA data contained in the
CASU QC tables.
The correlation coefficients have been multiplied by minus one, since
they would otherwise be negative due to the sky brightness being
in mag/arcsec$^2$.
Panel~(a) shows the Spearman rank order correlation coefficient $r_S$
while panel~(b) shows the linear Pearson correlation coefficient $r_P$.
The labels indicate the detector number.
\label{fig:nb118_daytime_test}
}
\end{figure}

To further investigate possible red-leaks in the 16 NB118 filters,
the ESO staff obtained a set of daytime test observations on 
2012 November 11 and 25 to perform a test suggested by W. Sutherland.
In addition to the normal screen flats, where a screen is illuminated by
a lamp, special screen flats were made with the lamp switched off
and the dome otherwise dark, i.e.\ with the illumination coming from
the thermal emission from the screen and the dome.
Both the ``lamp on'' and ``lamp off'' images were dark-subtracted
using dark frames taken the same day.
The median level in each detector was measured and converted to electrons
using the gain for each detector as determined from data taken the same day.
Finally, for each filter and detector, the ratio of the levels
in the ``lamp off'' and ``lamp on'' images was computed.
Over the 16 detectors, the median off/on ratios were
0.18\% in $J$, 2.5\% in NB118 and 48\% in $K_\mathrm{s}$.
Comparing $J$ with NB118 using these numbers suggest that the relative
importance of red-leaks in the NB118 filters is substantially larger
than in the $J$-band filters, but the stability of the ``off'' and ``on''
light sources between the exposures for the different filters is not
fully ensured.
Such problems are absent when comparing the off/on ratios for the
16 detectors of a given filter.
For $K_\mathrm{s}$ the range was small (47.1\%--49.6\%).
For $J$ the range was large (0.12\%--0.30\%), but the values themselves
are small.
For NB118 the range was large (1.4\%--4.1\%), 
with detectors 2 and 3 having the highest values.
These ratios are shown on the $y$-axes of Fig.~\ref{fig:nb118_daytime_test}.
The $x$-axes shows the correlation coefficients between
NB118 sky brightness in the given detector (using 1080 nighttime
observations taken over 1000 days) and ambient temperature,
as discussed above and as shown in Fig.~\ref{fig:sky_correlation}.
In other words, Fig.~\ref{fig:nb118_daytime_test} plots two different
measures of red-leak for the 16 NB118 filters against each other,
and they are indeed positively correlated.

It should be noted that the variation in red-leak strength within the
set of 16 filters derived here may partly reflect differences in
detector QE in the far red where the red-leaks may be located.
In the spare filters the red-leaks are located
near 2675\,nm, see Fig.~\ref{fig:sparefilters}.
We do not have the detector QE measurements for the individual detectors,
only for one detector, which is plotted in Fig.~\ref{fig:sparefilters}.
However, additional daytime test observations 
devised by V. Ivanov and obtained 2013 June 09
shed light on this question. In these tests, the filter wheel was
partially out of place in such a way that two NB118 filters partially
illuminated one detector. Data had to be saved directly from the
real-time display and only data for detector~2 were saved. 
One half of this detector was illuminated by filter~1, and the other half
by filter~2.
As in the tests described above, ``lamp off'' and ``lamp on'' images
were obtained and analysed.
In terms of the off/on ratio (hereafter $Q$), the results were as follows.
With the filter wheel in the normal position:
filter~1 on detector~1: $Q = 1.36\%$, and
filter~2 on detector~2: $Q = 3.23\%$, indicating that
filter~2 has 2.4 times more red-leak than filter~1
\emph{if the two detectors were identical}.
With the filter wheel in the special position:
filter~1 on detector~2: $Q = 2.51\%$, and
filter~2 on detector~2: $Q = 3.09\%$, indicating that
filter~2 has 1.2 times more red-leak than filter~1
\emph{when using the same detector (here detector~2)}.
These tests indicate that the relatively large red-leak seen for
NB118 filter~2 (when placed in front of detector~2 as in normal operation)
is mostly due to detector~2 having a relatively large QE at the location
of the red-leak(s) (whereever that may be).
We speculate that the same is the case for filter~3 and detector~3.
Detectors 2 and 3 are from an earlier batch than the other 14 detectors,
and detectors 2 and 3 also have a slightly worse AR (anti-reflective) coating
which can be seen in images of the
detectors\footnote{\url{http://www.vista.ac.uk/Images/FPA/IMGP4190.JPG}}
(W. Sutherland, priv.\ comm.\ 2013). It is therefore concievable that
detectors 2 and 3 could have a higher ratio of the QE at the red-leaks
of the NB118 filters relative to the QE at main passband of the NB118 filters
than the other 14 detectors.
As for the absolute QE in the $J$ and of $K_\mathrm{s}$ bands, our ``lamp on''
counts in electrons show that detectors 2 and 3 have a lower QE than
the other 14 detectors.

\subsection{Sky brightness near 1.19\,$\mu$m from other instruments}
\label{sec:NB118_sky_brightness_other_instruments}

As stated above, in the VISTA/VIRCAM NB118 data, the typical (over time)
sky brightness in the typical filter (over the 16 filters) is about
17.3\,mag/arcsec$^2$ (again, AB magnitudes are used throughout).
For the best filter, the value is about 17.7\,mag/arcsec$^2$,
and for the worst about 16.75\,mag/arcsec$^2$.
For reference, these filters probably have an average
central wavelength of about 1191\,nm (using the predicted values from
Sect.~\ref{sec:NB118_filters} and the inferred shift of unknown origin
of about 3.5--4\,nm from Sect.~\ref{sec:inferred_central_wavelength})
and an average FWHM of 12.3\,nm.
It is interesting to compare these VIRCAM NB118 sky brightness values to those
from other instruments having narrow-band filters near the 
1.19\,$\mu$m window.

For the KPNO~4m/NEWFIRM NB118 data of \citet{Ly_etal:2011},
the typical sky brightness is 18.16\,mag/arcsec$^2$ on average over the field,
with the best value obtained at the field centre (18.84\,mag/arcsec$^2$) and 
the worst at the field edge (about 17.9\,mag/arcsec$^2$)
(C. Ly, priv.\ comm.\ 2013).
The central wavelength of the filter changes with distance from the
field centre. At the field centre, the filter has
$\lambda_c$ = 1187.2\,nm and FWHM = 11.1\,nm (J. Lee, priv.\ comm.\ 2013).

For the Magellan/FourStar NB119 data of \citet{Lee_etal:2012},
the typical sky brightness is 18.0\,mag/arcsec$^2$
(J. Lee, priv.\ comm.\ 2013).
The filter has
$\lambda_c$ = 1190.8\,nm and FWHM = 13.7\,nm (J. Lee, priv.\ comm.\ 2013).
Note that this filter sometimes is called NB118 instead of NB119.

For the CFHT/WIRCam lowOH2 (NB118) data of Sobral et al.\ (in prep.),
the typical sky brightness is 18.5\,mag/arcsec$^2$
(D. Sobral, priv.\ comm.\ 2013).
The filter has
$\lambda_c$ = 1187\,nm and FWHM = 10\,nm (D. Sobral, priv.\ comm.\ 2013).

For the VLT/ISAAC NB119 data of \citet{Willis_Courbin:2005},
the typical sky brightness is $18.5\,\mathrm{mag}\,\mathrm{arcsec}^{-2}$.
The filter has
$\lambda_c$ = 1187\,nm and FWHM = 8.95\,nm.

The sky brightness in narrow-band filters is a combination of
inter-line sky continuum, line emission, and possibly thermal emission
entering via a red-leak. The brightness of the inter-line sky continuum
is debated but is likely low. For example,
\cite{Sullivan_Simcoe:2012} found the
inter-line sky continuum in the J band in dark time to be
19.55\,mag/arcsec$^2$ (with an uncertainty of about $\pm$0.2\,mag),
and \citet{Tilvi_etal:2010} found the sky brightness between the OH lines
near 1063\,nm (which is in the Y band)
to be about 21.2\,mag/arcsec$^2$.

\begin{figure*}[htbp]
\centering
\includegraphics[width=0.90\textwidth,bb=-54 191 652 609]{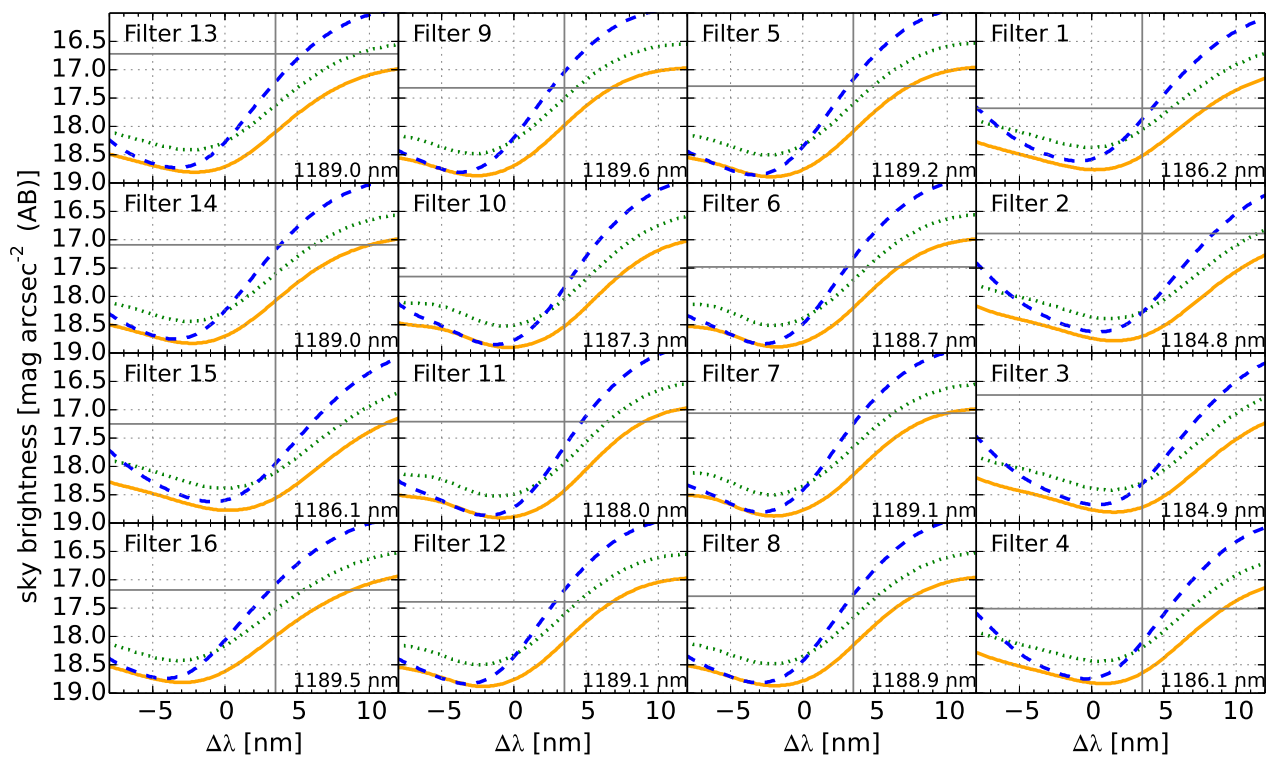}
\caption[]{%
Predicted sky brightness for the 16 NB118 filters as function of a
hypothetical filter wavelength shift $\Delta\lambda$ with respect to the filter
curves as predicted based on laboratory measurements.
The results from 3 models of the sky emission spectrum are shown:
solid orange line: model~1 (Gemini Observatory, theoretical),
dotted green line: model~2 \citep{Sullivan_Simcoe:2012},
dashed blue line:  model~3 (see text).
The predictions are based on sky continuum and emission lines near the
central passband of the filters; possible red-leaks are not included.
The horizontal grey line marks a representative value for the observed
NB118 sky brightness in the given filter.
The vertical grey line marks the wavelength shift to the red of about
3.5\,nm that we infer from NB118 observations of emission line objects with
spectroscopic redshifts.
The mean filter wavelength $\lambda_0$ (Eq.~\ref{eq:meanwave})
as predicted based on laboratory measurements is given on the panel. 
\label{fig:skybrightness_theo}
}
\end{figure*}

\section{Predicted sky brightness}
\label{sec:predicted_sky_brightness}

In this section we will calculate the predicted sky brightness for each of
the 16 NB118 filters for a given model of the sky spectrum. We will do this
for the filter curves as predicted based on laboratory measurements,
and for these filter curves shifted in small steps in wavelength
$\Delta\lambda$.
The aim is to evaluate the influence of the wavelength shift of about
3.5--4\,nm to the red that we have inferred using emission line objects
with spectroscopic redshifts (Sect.~\ref{sec:inferred_central_wavelength}).
The aim is also to evaluate the amount of sky brightness caused by red-leaks
--- our red-leak indicators (Sect.~\ref{sec:NB118_sky_brightness_all_VISTA})
mainly provide information about the relative importance of red-leaks within
the set of the 16 NB118 filters, not the absolute sky brightness due to
red-leaks.

Here we will use the following 3 models for the sky background.
Model~1 is the theoretical sky spectrum provided by Gemini Observatory,
as described in Sect.~\ref{sec:NB118_filters}.
Model~2 is the observed sky spectrum from \citet{Sullivan_Simcoe:2012}.
Model~3, like the other two models, contains a combination of sky lines
and continuum, although not directly in the form of a spectrum.
The sky lines are taken from the list of wavelengths and intensities of
sky emission lines as compiled from the literature
\citep{Maihara_etal:1993,Ramsay_etal:1992,Oliva_Origlia:1992,Abrams_etal:1994}
by the DAzLE team \citep[e.g.,][]{Horton_etal:2004,McMahon_etal:2008}
and made available on the DAzLE
web site\footnote{\url{http://www.ast.cam.ac.uk/~optics/dazle/sky-data/oh_sky_lines.txt}},
and the continuum emission of
$600\,\mathrm{photons}\,\,\mathrm{s}^{-1}\,\mathrm{m}^{-2}\,\mathrm{arcsec}^{-1}\,\mu\mathrm{m}^{-1}$
is taken from \citet{Maihara_etal:1993}.

In our calculation we use the filter curves as predicted based on
laboratory measurements, similar to those in the fourth row of
Fig.~\ref{fig:filtercurves}, except that atmospheric absorption is
not included.
In the calculation 
we assume that the shape of the bandpasses is preserved and only a shift
of the complete filter in wavelength is applied.  

It is important to note that since the filter curves for the 16 NB118 filters
installed in VIRCAM were only measured in a wavelength range near the
central passband, our calculations of sky brightness do not include
possible red-leaks.

The result in terms of sky brightness versus $\Delta\lambda$
is given in Fig.~\ref{fig:skybrightness_theo}.
The 3 models give somewhat different results.
In terms of the effect of the filters probably being about
3.5\,nm redder than
predicted from the laboratory measurements,
i.e.\ the difference in sky brightness between
$\Delta\lambda = 3.5\,\textrm{nm}$ and
$\Delta\lambda = 0\,\textrm{nm}$,
the effect varies with filter and model.
For model~1, the average is $-0.46$\,mag (range: $-0.04$ to $-0.71$).
For model~2, the average is $-0.51$\,mag (range: $-0.09$ to $-0.75$).
For model~3, the average is $-0.93$\,mag (range: $-0.34$ to $-1.19$).
The sign is such that the sky always is brighter when the filters
are shifted 3.5\,nm to the red.

In terms of the absolute sky brightness at
$\Delta\lambda = 3.5\,\textrm{nm}$, the models vary a lot.
On the figure the grey horizontal line marks the observed median (over time)
NB118 sky brightness as measured by the CASU QC for UltraVISTA data
(Sect.~\ref{sec:NB118_sky_brightness_all_VISTA});
the values for the GTO data would be about 0.1\,mag brighter.
The difference between the observed and the predicted sky brightness
can be interpreted as the contribution due to a red-leak.
For models~1 and 2, most filters seem to have a contribution from red-leaks.
For model~3 the sky brightness for most filters can be explained without
red-leaks.
All models predicted a much darker sky for filters~2 and 3 than observed,
and these two filters (combined with their respective detectors)
are also those that show the strongest signs of
red-leaks (Sect.~\ref{sec:NB118_sky_brightness_all_VISTA}).
We note that our (rather uncertain) estimate of the red-leak contribution for
the spare filters was typically only $\sim35\%$ = 0.3\,mag
(Sect.~\ref{sec:NB118_filters}).

To make further progress, we should base our sky spectrum on flux calibrated
NIR spectra from instruments at Paranal (e.g.\ X-shooter) located just
1.5\,km away from VISTA and taken simultaneously with the VISTA NB118 data in
question. We are considering this for a future paper.

\section{Summary and Conclusions}
\label{sec:summary}

We have designed and had installed a set of 16 narrow band
filters centred near 1.19$\,\mu$m in the ESO/VISTA instrument
VIRCAM\@. 
We here report on 3 nights of
GTO observations with this filter set. In particular
we describe its performance in terms of sky noise,
throughput and passband and we describe in detail what
we have found to be the best observing and reduction
strategies with this filter set.

The filters are designed to fit into a spectral region
with few airglow lines and therefore relatively low sky
background. The central wavelength corresponds to
redshifts of $z \approx$ 0.8, 1.4, 2.2 and 8.8 for emission lines
H$\alpha$, H$\beta$+[\ion{O}{iii}], [\ion{O}{ii}] and Ly$\alpha$,
respectively, which combined with the large
area covered by VIRCAM in a single exposure makes it
a very powerful facility for emission line surveys in the near-infrared.

We found that persistence was an important issue in our
data and that persistence from bright objects could be
detected not only in the following image but also in the
one following that. The strength of the persistence is
not the same for all detectors and we therefore designed
a general process for masking which works optimally for
all 16 detectors after we determine a number of detector-dependent
parameters. We describe this process in
detail since it will be of importance for other
projects using VISTA data, but we also point out that
the problem in our data was enhanced by our choice of
OB nesting and make specific recommendations for which
nesting to use for this kind of observations.

We found that narrow-band data from different detectors show a
large variation (factor of up to 2.43) in sky background
and conclude that this is most likely related to slight
differences in the passbands and/or red-leaks of the
narrow-band filters. This produces variations in the depth
within the combined image.

From a cross correlation of catalogues of spectroscopic
redshifts in the field and a preliminary catalogue of
emission line objects selected in our narrow-band stack we confirm
that we are able to identify emission line galaxies broadly as expected,
but the passband is shifted towards slightly higher
redshifts. The shift is about 3.5--4\,nm towards the red;
the origin of the shift is unknown.

The full scientific exploitation of the GTO data with
a complete catalogue of extracted emission line objects
to well defined detection limits will be presented
elsewhere, but we did extract a preliminary
list of emission line objects which was necessary in
order to characterise the performance of the filters.

As mentioned above we have cross correlated this list
with known spectroscopic redshifts and we find that we
have identified objects at a large number of redshifts
with a corresponding range of different emission lines 
in the narrow filters. In addition to the
expected strong lines 
([\ion{S}{ii}]6718,6733, H$\alpha$+[\ion{N}{ii}], H$\beta$+[\ion{O}{iii}]
and [\ion{O}{ii}]3727,3730)
we detect the [\ion{S}{iii}]9533 line not usually targeted in
emission-line surveys.
We also find one likely \ion{Mg}{ii}2796,2804 emitter,
which is classified as a broad-line AGN by zCOSMOS\@.

\section*{Acknowledgements}

We thank the ESO Paranal staff, in particular
Thomas Szeifert, Carlos La Fuente, Valentin Ivanov and Andres Parraguez,
for assistance during the observations.
We thank the CASU staff,
in particular Jim Lewis, Mike Irwin and Eduardo Gonzalez-Solares,
for performing the initial data processing.
We thank Olivier H{\'e}rent, Patrick Hudelot and Yuliana Goranova for
help with the data processing at TERAPIX\@.
We thank
Will Sutherland,
Jim Emerson,
Gavin Dalton,
Michael I. Andersen,
Sangeeta Malhotra,
James Rhoads,
Chun Ly,
Janice Lee,
Marina Rejkuba,
John Moustakes,
and
Lise Christensen
for discussions.
We thank Marcus Wallace from NDC for information about the filters.
We thank the referee for a constructive report that prompted us to
improve both the reduction and parts of the presentation.
BMJ thanks ESO for hospitality during a 3 month visit.
We gratefully acknowledge funding for purchasing the NB118 filters from
the Dark Cosmology Centre and from IDA (Instrumentcenter for Dansk Astrofysik).
%
This paper makes use of observations processed by the Cambridge Astronomy
Survey Unit (CASU) at the Institute of Astronomy, University of Cambridge.
This work is based on data products produced at the TERAPIX data center
located at the Institut d'Astrophysique de Paris.
Based in part on data products produced by TERAPIX and
the Cambridge Astronomy Survey Unit on behalf of the
UltraVISTA consortium.
Based in part on zCOSMOS observations made with ESO Telescopes at the La Silla
or Paranal Observatories under programme ID 175.A-0839.
BMJ and JPUF acknowledge support from the ERC-StG grant EGGS-278202.
The Dark Cosmology Centre is funded by the Danish National Research
Foundation.
JSD acknowledges the support of the European Research
Council via the award of an Advanced Grant, and
the support of the Royal Society via a Wolfson Research Merit Award.

\bibliography{papers_cited_by_milvang}

\begin{appendix}

\section{Creation of persistence masks}
\label{ap:persistence}

\begin{figure*}[htbp]
\centering
\includegraphics[width=0.60\textwidth]{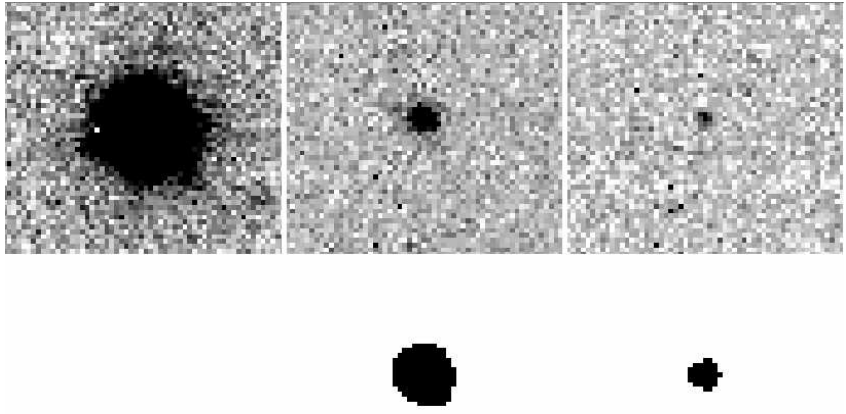}
\caption[]{%
Illustration of persistence in individual VIRCAM images.
The top row shows a subsection of 3 NB118 images (DIT = 280 sec, NDIT = 1)
taken right after each other. The data are from detector 3, and the 3
subsections correspond to the same pixels on the detector. In between each
image the telescope has moved, so the subsections correspond to different
parts of the sky. The width of the section is 19$''$, and the pixel scale
is the native one of $0.34'\,\mathrm{px}^{-1}$.
In the first image a star of magnitude 
$J_\mathrm{Vega} = 11.1$ (2MASS)
is seen. In the subsequent two images, there is  a persistent image
at the same detector coordinates as the star in the first image.
The bottom row shows the persistence masks created by our algorithm for each image, see text.
\label{fig:persistence_indiv_images}
}
\end{figure*}

Most infrared detectors suffer from persistence, i.e.\ a sufficiently bright
source will produce a signal in one or more subsequent exposures.
The 16 VIRCAM detectors, of type Raytheon VIRGO $2048 \times 2048$ HgCdTe on
CdZnTe substrate \citep[e.g.][]{Bezawada_etal:2004,Bezawada_Ives:2006},
are reported to have low persistence\footnote{\url{http://casu.ast.cam.ac.uk/surveys-projects/vista/technical/persistence}},
but the effect is nevertheless easily detectable and can for certain
observing patterns (i.e.\ an unfortunate choice of the so-called nesting)
cause severe problems, as we will show below.
We find that the object causing the persistence (typically a star) does not
have to be extremely bright: in our stack we would see persistence from
stars as faint as $J_\mathrm{Vega} \approx 15$ ($J_\mathrm{AB} \approx 16$)
had we not applied our persistence masking. Furthermore,
for some of the detectors the star does not even have to be saturated
to cause persistence.

At the level of the individual images, i.e.\ before stacking,
persistence will be present in most VIRCAM images (regardless of filter).
This is illustrated in the top row of Fig.~\ref{fig:persistence_indiv_images},
where we show a small section of 3 images taken after each other. The shown
section is the same on the detector but different on the sky, as the telescope
was moved between each image. The first image contains a somewhat bright star
($J_\mathrm{Vega} \approx 11$).
The two subsequent images do not contain any astronomical objects, and
the source that is seen is the so-called persistent image of the star,
i.e.\ a fake source.

\begin{figure*}[htbp]
\centering
\includegraphics[width=0.60\textwidth,bb=40 244 555 562,clip=]{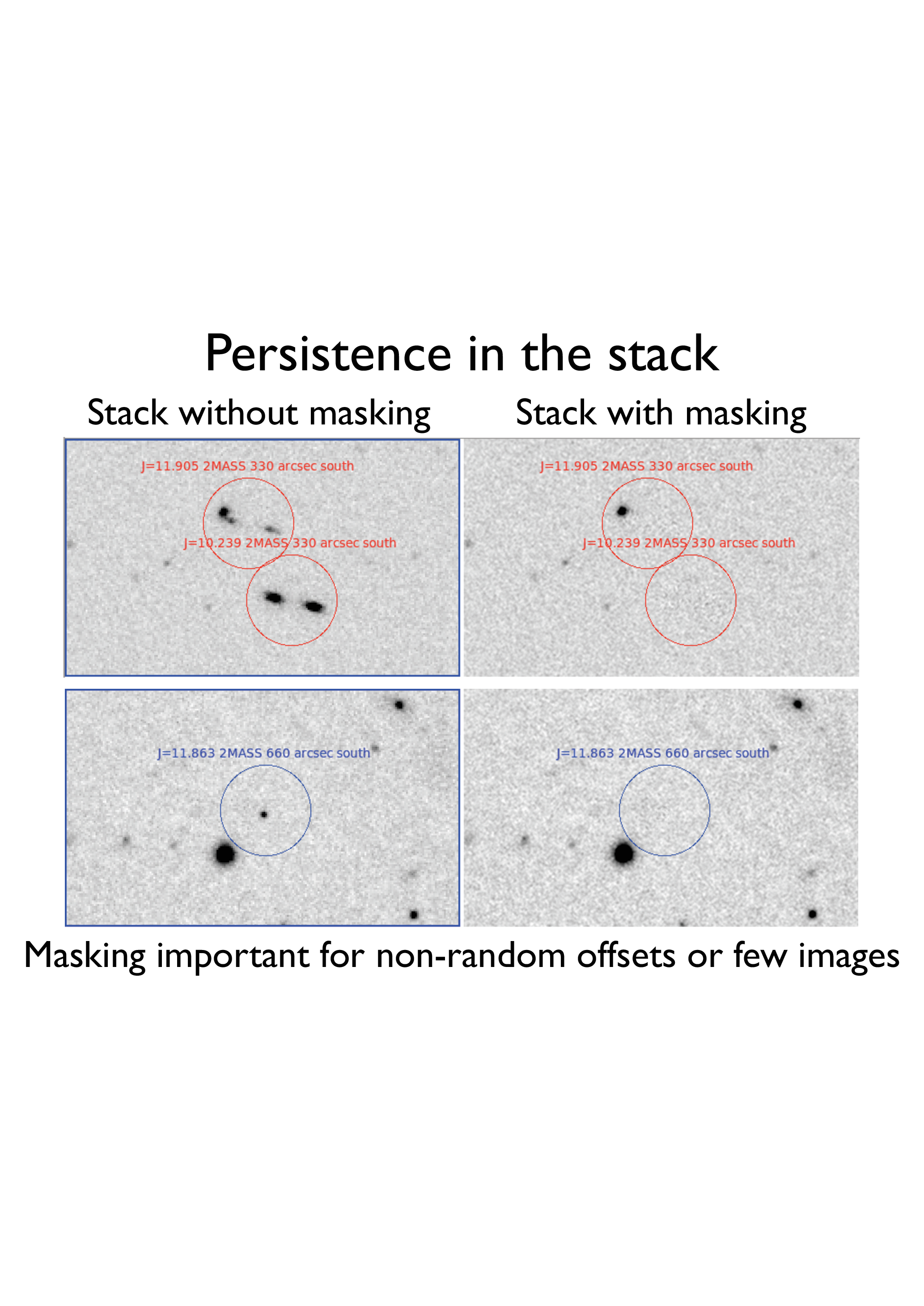}
\caption[]{%
Illustration of persistence in the stack of our data.
Left: stack made without masking the persistent images (i.e.\ fake sources)
in the individual images.
Right: stack made with masking.
The circles mark the predicted locations of persistent images in the stack,
see text. The circles are labelled with the $J$-band Vega magnitude of the
star that caused the persistence in question.
The diameter of the circles is 15$''$. North is up and east is left
\label{fig:persistence_stack}
}
\end{figure*}

Whether persistent images will be present in the stack depends on
the number of individual images being stacked and on the observing pattern
(i.e.\ on the type of offsets/jitters) used to obtain these images:
\begin{itemize}
\item
If many individual images are stacked and if random jitters have been used
between each image,
then the persistent images will not ``stack up'', and sigma clipping will
effectively remove these.
\item
If only a few individual images are stacked, then sigma clipping will not
be able to (fully) remove the persistent images.
\item
If an unfortunate observing pattern is used, as is the case here
(but which may be rare),
then stacking many images does not help, as the the persistent images will
``stack up'' and sigma clipping will not remove them.
\end{itemize}
The latter situation is illustrated in Fig.~\ref{fig:persistence_stack}.
The two panels on the left hand side show a stack of our NB118 data made using
sigma clipping, but without applying the masking that we will describe below.
The circles are located at the predicted positions of persistent images 
(fake sources) in the stack. 
The two panels on the right hand side show the stack in which the
persistent images in each individual image have been masked, and it is seen
that the masking removes the fake objects, while leaving real objects intact.
Note that the masking operates on the individual images, not on the stack
directly.

The main point of Fig.~\ref{fig:persistence_stack} is that
without masking we would have persistent images in our stack,
and these could be mistaken for emission line objects, since they would
mostly be present in the NB118 stack and less so in the $J$-band stack where
most of the data come from UltraVISTA where random offsets between each
image were used.

The details of Fig.~\ref{fig:persistence_stack} are not so important,
but can nevertheless be described as follows.
As described in Sect.~\ref{sec:obs}, we used an observing pattern in which
the offset between the 3 pawprint positions (named paw6, paw5 and paw4,
see Fig.~\ref{fig:pawprint}) was always 5.5$'$ in Dec,
and in which a random offset was only applied after each set of 3 pawprints.
Recalling from the top row of Fig.~\ref{fig:persistence_indiv_images} that persistent
images can sometimes be seen both in the first and the second image after a
given image in which a bright star was located, we expect persistent images to
be present in the stack at locations 5.5$'$ and 11$'$ north of bright stars.
The circles in Fig.~\ref{fig:persistence_stack} correspond to modified versions
of the 2MASS catalogue 
in which 5.5$'$ (top panel) or 11$'$ (bottom panel)
has been added to the Dec.
In the 5.5$'$ case (top panel) a given bright star gives rise to two
persistent images separated by about 7$''$
(the circles in the figure have a diameter of 15$''$)
mostly along the east-west direction.
One comes from the star being in paw6 and the persistent image being in paw5,
and the other comes from the same star being in paw5 and the persistent image
being in paw4.
The telescope movements paw6$\rightarrow$paw5 and paw5$\rightarrow$paw4
must have had a small (about 3.5$''$) and different component in RA (in
addition to the 5.5$'$ in Dec) giving rise to the double appearance of these
fake sources. The elongation of sources are due to the RA component of the
offset being slightly different for each of the 6 nights from which we have
data. In summary, the morphology of these fake sources is fully understood.
It is noteworthy that the fake sources in the 11$'$ case (bottom panel)
look like real objects and could therefore not have been excluded based on
their morphology.

We will now describe the algorithm we have developed for masking
the persistent images in the individual images.
The algorithm uses a set of parameters, which we derived from a set of
VISTA data with different background levels in order to make the
algorithm more generally applicable.
Specifically, we used data from this work, meaning
NB118 (DIT = 280\,sec, NDIT = 1) and $J$ (DIT = 30\,sec, NDIT = 4),
and NB118 data from the VISTA SV programme\footnote{%
We used data in the field of NGC 253, ESO ID 60.A-9285(A);
the VISTA SV is described in \citet{Arnaboldi_etal:2010} and at
\url{http://www.eso.org/sci/activities/vistasv/VISTA_SV.html}},
with DIT ranging from 270\,sec to 860\,sec, all with NDIT = 1.
The SV data were also chosen since some of these OBs also had
an unfortunate nesting,
meaning that masking of persistence is needed to make a clean stack.

We proceeded as follows.
We first created a list of images taken right after each other.
These images are the individual images as reduced by
CASU (Sect.~\ref{sec:CASU}).
We focused on a given image somewhere in the list, called image~0,
and we then investigated how the pixel values in the 3 subsequent images
(called image~1, image~2 and image~3) were correlated with the pixel values
in image~0.
When analysing each image we subtracted the median value to get a background
that is approximately zero. We restricted our analysis to manually
selected subsections that contained a star in image~0 and no astronomical
objects in the subsequent images.
For each pixel we plotted the value in image~1 (or image~2 or image~3)
against the value in image~0.
Since the telescope moved between the two images in question,
the values should be uncorrelated. However, due to the persistence this is not
the case; rather, a positive correction is always seen in the image~1 case,
and for some detectors also in the image~2 case. No correlations were seen
in the image~3 case.

\begin{figure*}[htbp]
\centering
\includegraphics[width=0.80\textwidth,bb=11 252 570 571]{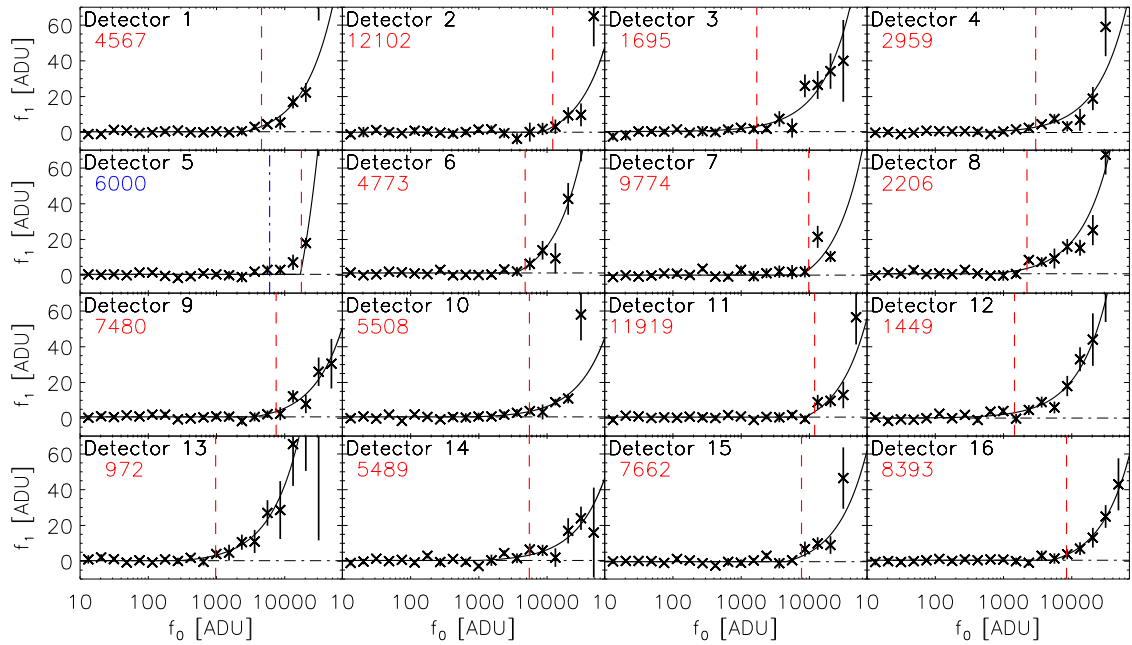}
\caption[]{%
Persistence in the 16 VIRCAM detectors.
The figure is based on a number of image pairs taken right after each other,
called image~0 and image~1.
Only subsections with a somewhat bright star in image~0 and
no astronomical object in image~1 were analysed
(the telescope moved in between the two images).
In each image the median value over the image, i.e.\ the background,
was subtracted. For each pixel, the fluxes $f_0$ and $f_1$ in the two images
were measured. The figure shows a plot of these, binned in $f_0$.
The persistence (i.e.\ detector memory effect) is seen by the fact that
$f_1$ becomes positive above a certain threshold in $f_0$.
The black solid line is a fit to the data
(Eq.~\ref{eq:persistence_fitting_function_1_vs_0}).
The red dashed line is the threshold in $f_0$ at which the fit reaches
a value of 3\,ADU above the background, and this is used in our
masking algorithm as the threshold for masking a given pixel (i.e.\ for
excluding it when stacking the data). For detector 5 a lower threshold
was manually set, as indicated by the blue dot-dashed line.
The used thresholds are given on the panels and listed in
Table~\ref{tab:persistence}.
The $f_0$ axis is logarithmic and runs from 10 to 70,000.
The units are gain-normalised ADU, as used in the individual images 
reduced by CASU.
\label{fig:persistence_image1_vs_image0}
}
\end{figure*}

\begin{figure*}[htbp]
\centering
\includegraphics[width=0.80\textwidth,bb=11 252 570 571]{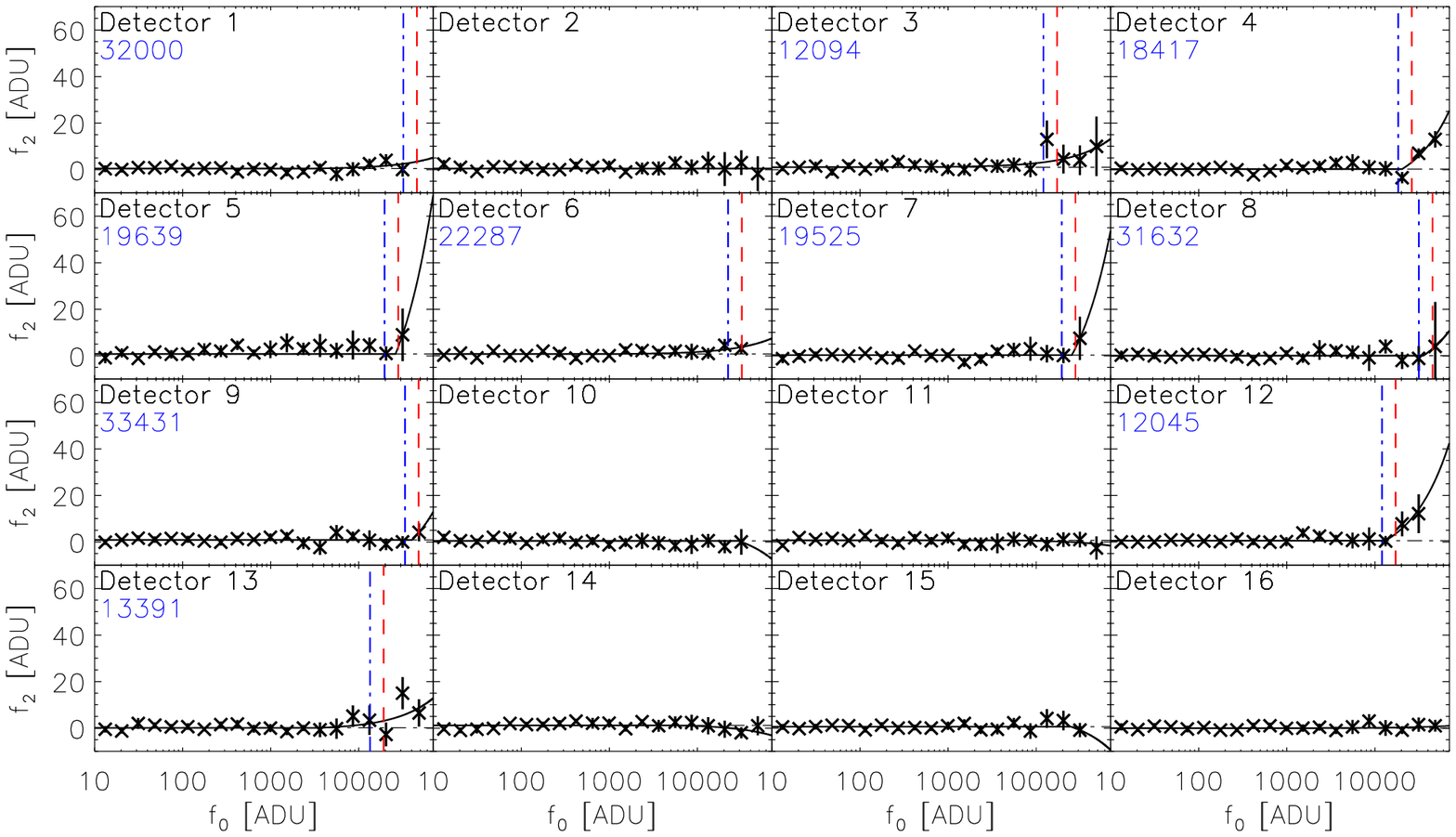}
\caption[]{%
Persistence in the 16 VIRCAM detectors.
This figure shows the persistence in image~2, i.e.\ the second image
after a given image (image~0) containing somewhat bright pixels.
In most aspects this figure is as Fig.~\ref{fig:persistence_image1_vs_image0}.
One difference is that the blue dot-dashed lines show the adopted
masking threshold (given on the panels and in Table~\ref{tab:persistence}),
which is 30\% lower than the threshold derived from the fit (marked by the
red dashed lines). Some detectors do not show detectable persistence in
image~2, and for these no threshold was derived.
\label{fig:persistence_image2_vs_image0}
}
\end{figure*}

The image~1 vs image~0 plots for the 16 detectors
are shown in Fig.~\ref{fig:persistence_image1_vs_image0}, with
the background-subtracted level in image~1 (denoted $f_1$) on the y-axis and
the background-subtracted level in image~0 (denoted $f_0$) on the x-axis.
To reduce the noise we have binned the data in $f_0$, and we have
combined the different datasets that we analysed (see below).
The persistence effect is clearly seen for all 16 detectors:
when a certain threshold in $f_0$ is reached, $f_1$ becomes non-zero and
increases with $f_0$.

The datasets that we analysed consisted of 7 sets of images,
which spanned the two filters (NB118 and $J$) and a range of DIT values.
In each set of images we analysed two subsections.
We also made versions of Fig.~\ref{fig:persistence_image1_vs_image0} in which the
different filters and DIT values were plotted separately, and no
obvious dependence on filter or DIT were seen.
The simple procedure of subtracting the median (i.e.\ the background)
allowed datasets with different background levels to coincide in these plots,
greatly simplifying the analysis.

The data points in Fig.~\ref{fig:persistence_image1_vs_image0} indicate that we could
in principle correct for the persistence by using $f_0$ to predict how
much persistence signal $f_1$ should be subtracted. However, we will do
something simpler and more robust: we will determine a threshold in $f_0$
above which the pixel in image~1 should be considered unreliable and therefore
be masked (i.e.\ be given zero weight when stacking the data).
To determine this threshold for each detector, we fit functions 
with 3 free parameters ($k_1$, $a_1$ and $b_1$) of the form
\begin{equation}
f_1(f_0) =
  \begin{cases}
    k_1                     & \mbox{if\,\,} f_0 \le a_1 \\
    k_1 + b_1\cdot(f_0-a_1) & \mbox{if\,\,} f_0 > a_1
  \end{cases}
\label{eq:persistence_fitting_function_1_vs_0} 
\end{equation} 
to the data points in Fig.~\ref{fig:persistence_image1_vs_image0}.
These fitted functions are shown as the black, solid lines in
Fig.~\ref{fig:persistence_image1_vs_image0}; note that the fit is (piecewise) linear,
but the figure uses a logarithmic x-axis.
We then define the threshold (called ``threshold 1'', as listed in
Table~\ref{tab:persistence}) above which we need to apply the masking
as the $f_0$ value at which the fit reaches 3\,ADU above the constant level
$k_1$ (which is essentially zero).
This threshold is shown as the red, dashed, vertical line in
Fig.~\ref{fig:persistence_image1_vs_image0}.
For detector 5 the derived threshold seemed too high given the data points,
and we manually set it to a lower value, as also shown on the figure.

The level of 3\,ADU was chosen since for our data, a persistence signal
below this level would lead to a signal of less than 1$\sigma$ of the
noise in the stack.

\begin{table}
\caption{Parameters used in the persistence masking algorithm
\label{tab:persistence}}
\centering
\begin{tabular}{rrrr}
\hline
\hline
Detector & threshold~1 & threshold~2        & saturation \\ \hline
  1      &  4567       &  32000\tablefootmark{a} & 30000 \\
  2      & 12102       &    NaN\tablefootmark{a} & 39300 \\
  3      &  1695       &  12094                  & 39200 \\
  4      &  2959       &  18417                  & 38900 \\
  5      &  6000       &  19639                  & 31000 \\
  6      &  4773       &  22287                  & 32200 \\
  7      &  9774       &  19525                  & 30000 \\
  8      &  2206       &  31632                  & 39300 \\
  9      &  7480       &  33431                  & 40800 \\
 10      &  5508       &    NaN                  & 34000 \\
 11      & 11919       &    NaN                  & 40200 \\
 12      &  1449       &  12045                  & 32000 \\
 13      &   972       &  13391                  & 41400 \\
 14      &  5489       &    NaN                  & 40200 \\
 15      &  7662       &    NaN                  & 34000 \\
 16      &  8393       &    NaN                  & 41800 \\
\hline
\end{tabular}
\tablefoot{%
The units are gain-normalised ADU, as used in the individual images 
reduced by CASU, see Sect.~\ref{sec:CASU}.
``NaN'' (Not a Number) indicates that no threshold was set.
\tablefoottext{a}{This threshold was subsequently found to be too high}
}
\end{table}

The case of image~2 vs image~0 is shown in Fig.~\ref{fig:persistence_image2_vs_image0}.
For some detectors, e.g.\ detector~12, persistence is still seen,
but only occurring above a higher threshold in $f_0$.
For each detector, a function of the same functional form as before,
\begin{equation}
f_2(f_0) =
  \begin{cases}
    k_2                     & \mbox{if\,\,} f_0 \le a_2 \\
    k_2 + b_2\cdot(f_0-a_2) & \mbox{if\,\,} f_0 > a_2
  \end{cases}
\label{eq:persistence_fitting_function_2_vs_0} 
\end{equation} 
was fitted to the data points in Fig.~\ref{fig:persistence_image2_vs_image0};
the fits are shown on the figure.
To get the masking threshold we first calculated the level at which the
fit reached 3\,ADU above the constant level; this threshold is shown as
the red, dashed, vertical line in the figure. Experimentation showed that
slightly more masking was needed, so we multiplied the threshold by 0.7,
shown by the blue, dot-dashed, vertical line in the figure.
These latter thresholds are listed as ``threshold 2'' in
Table~\ref{tab:persistence}.
We note that the determination of these threshold~2 values has
much higher uncertainty than those for image~1, since the signal is lower
(cf.\ Fig~\ref{fig:persistence_image2_vs_image0}), which in turn is due to only
a few pixels being bright enough to lead to a persistence signal in
the second subsequent image.    
This warrants lowering the thresholds by 30\%.
For some detectors no persistence was seen in image~2, and for these
no threshold~2 value was set (cf.\ Table~\ref{tab:persistence}).
It turned out that the stack made with the masking based on the
thresholds in Table~\ref{tab:persistence}
contained a few (4 in total) surviving ``image~2'' persistent images
from detectors 1 and 2. Therefore, the used threshold~2 values for these
two detectors should have been set lower; this has been indicated in
Table~\ref{tab:persistence}.

Also the image~3 vs image~0 case was investigated, but no persistence
was detected.

The analysis above shows that persistence masks can be generated simply by
measuring the background-subtracted flux $f_0$ in each pixel in image~0,
and then flag the given pixel in the masks for image~1 and/or image~2
if $f_0$ is above the thresholds listed in Table~\ref{tab:persistence}.
One problem remains, related to saturated stars. These stars
intrinsically have high fluxes in their centres, but in the images
delivered by VIRCAM these stars can have low values in their centres.
Such donut shaped stars with a dip or hole in the centre
are due to the CDS readout mode where each image is actually
the difference of a long and a short exposure, see
Sect.~\ref{sec:Correlated_Double_Sample} and the CASU web site\footnote{%
\url{http://casu.ast.cam.ac.uk/surveys-projects/vista/technical/known-issues}}.
In an image containing stars of a range of magnitudes, this phenomenon leads
to the following effect. Stars with peak level below the saturation threshold
have the shape of the expected point spread function. When looking at
increasingly brighter stars, one first sees stars with flat centres, then
stars with a slight dip at the centre, and finally stars with a dip that
reaches down to zero. We found that the centres of these objects with
dips/holes still cause persistence in subsequent images, even if when their
flux values is below the thresholds of Table~\ref{tab:persistence}.
This has to be taken into account in any algorithm to address the persistence
problem.

We used the above findings to generate persistence mask for each individual
image in the following manner. First, we generated versions of the images
in which the holes in the centre of bright stars were set to a high value.
We did this by first identifying groups of connected pixels having a
flux above 80\% of the saturation level listed in Table~\ref{tab:persistence}.
For a star with a hole in the middle, such a group of pixels would be a ring
around the centre of the star. Within each group of connected, high-valued
pixels, a line was drawn between all the pixels with the same x-coordinate,
and similarly for y, and all the pixels that these lines passed,
including pixels with low values such as those forming the hole/dip,
were set to a high value (specifically 1.1 times the saturation level,
but the exact value is not important).

The mask were created from these modified images by marking all pixel in
which the median-subtracted flux in the preceding image exceeds
threshold~1, and in which the median-subtracted flux in the 
pre-preceding image exceeds threshold~2. These pixels were considered to be
affected by persistence and therefore ignored (i.e.\ given zero weight)
when stacking the images.  
Examples of the computed persistence masks are shown
in the bottom row of Fig.~\ref{fig:persistence_indiv_images}.
Typically, 0.1\% of the pixels in a given individual image are masked.
There is some variation from detector to detector, given that some detectors
(e.g.\ detectors 3, 12 and 13), have stronger persistence
(i.e.\ need less bright pixels to cause persistence) than others,
cf.\ Figs.~\ref{fig:persistence_image1_vs_image0} and
\ref{fig:persistence_image2_vs_image0} and
Table~\ref{tab:persistence}.

The algorithm works both for NDIT = 1 and NDIT $>$ 1 images.
In the case of NDIT $>$ 1, the FITS file contains the sum of the
NDIT sub-exposures. The pixel values simply need to be divided by NDIT\@.
Care has to be taken that CASU sets the BSCALE header keyword to
1/NDIT, which makes some stronomical software (but not all) silently scale
the data by this factor.

An IDL implementation of the algorithm is available upon request.

Our persistence masking algorithm was also used in the stacking of the
UltraVISTA NB118 season 1 data \citep{McCracken_etal:2012}.

\end{appendix}

\end{document}